\documentclass[useAMS,usenatbib,articles] {mn2e}
\usepackage{epsfig}
\usepackage{graphicx}
\usepackage{lscape}
\usepackage{amssymb} 

\title[Origin of the IMF of stellar clusters]{The IMF of stellar clusters: effects of accretion and feedback}

\author[S. Dib et al.]{Sami Dib$^{1}$\thanks{E-mail: sami.dib@cea.fr}, Mohsen Shadmehri$^{2}$, Paolo Padoan$^{3}$, Maheswar, G.$^{4}$, D. K. Ojha$^{5}$, 
\newauthor Fazeleh Khajenabi$^{6}$\\                           
$^{1}$Service d'Astrophysique, DSM/Irfu, CEA/Saclay, F-91191, Gif-sur-Yvette, Cedex, France\\
$^{2}$Department of Mathematical Physics, National University Ireland, Co kildare, Maynooth, Ireland \\
$^{3}$ICREA-ICC, University of Barcelona, Spain\\ 
$^{4}$Aryabhatta Research Institute of Observational Sciences, Manora Peak, Nainital 263129, India \\
$^{5}$Tata Institute for Fundamental Research (TIFR), Homi Bhabha Road, Mumbai-400005, India\\ 
$^{6}$School of Physics, University College Dublin, Belfield, Dublin 4, Ireland\\ }

\begin{document}
\maketitle

\date{Accepted XXX. Received XXX}

\pagerange{\pageref{firstpage}--\pageref{lastpage}}
\pubyear{2009}
\label{firstpage}

\begin{abstract} 
We have developed a model which describes the co-evolution of the mass function of dense gravitationally bound cores and of the stellar mass function in a protocluster clump. In the model, dense cores are injected, at a uniform rate, at different locations in the clump and evolve under the effect of gas accretion. Gas accretion onto the cores follows a time-dependent accretion rate that describes accretion in a turbulent medium. Once the accretion timescales of cores of a given age, of a given mass, and located at a given distance from the protocluster clumps center exceed their contraction timescales, they are turned into stars. The stellar initial mass function (IMF) is thus built up from successive generations of cores that undergo this accretion-collapse process. We also include the effect of feedback by the newly formed massive stars through their stellar winds. A fraction of the wind's energy is assumed to counter gravity and disperse the gas from the protocluster and as a consequence, quench further star formation. The latter effect sets the final IMF of the cluster. We apply our model to a clump that is expected to resemble the progenitor clump of the Orion Nebula Cluster (ONC). The ONC is the only known cluster for which a well determined IMF exists for masses ranging from the sub-stellar regime to very massive stars. Our model is able to reproduce both the shape and normalization of the ONC's IMF and the mass function of dense submillimeter cores in Orion. The complex features of the ONC's present day IMF, namely, a shallow slope in the mass range $\sim [0.3-2.5]$ M$_{\odot}$, a steeper slope in the mass range  $\sim [2.5-12]$ M$_{\odot}$, and a nearly flat tail at the high mass end are reproduced. The model predicts a 'rapid' star formation process with an age spread for the stars of $2.3~\times 10^{5}$ yr which is consistent with the fact that 80 percent of the ONC's stars have ages of $\lesssim 0.3$ Myr. The model also predicts a primordial mass segregation with the most massive stars being born in the region between 2 and 4 times the core radius of the cluster. In parallel, the model also reproduces, at the time the IMF is set and star formation quenched, the mass distribution of dense cores in the Orion star forming complex. We study the effects of varying some of the model parameters on the resulting IMF and we show that the IMF of stellar clusters is expected to show significant variations, provided variations in the clumps and cores physical properties exist.
\end{abstract} 


\section{INTRODUCTION}\label{intro}

\subsection{ORIGIN OF THE IMF}\label{origin}
The origin of the stellar initial mass function (IMF) and its potential universality are among some of the most challenging issues in modern astrophysics. Salpeter (1955) made the first observational measurement of the IMF for field stars whose masses are $\gtrsim 0.3$ M$_{\odot}$. He obtained an IMF that follows a power law of the form $dN/dM=M^{\psi \sim -2.35}$ (or $dN/dlog M=M^{\Gamma \sim -1.35}$ ). Subsequent work by several authors showed that the IMF possesses a slope of $\psi_{l} \sim -0.1$ at low masses (i.e., M $\lesssim 0.1$ M$_{\odot}$) mediated by a plateau-like regime with a shallow slope $\psi_{i} \sim -0.3$ in the mass range $0.1-0.5$ M$_{\odot}$ and a steeper slope, $\psi_{h} $, at larger masses. Whereas Miller \& Scalo (1979), Scalo (1998), and Chabrier (2003) argued for a lognormal form for the IMF, Kroupa (2002) proposed  that the field stars IMF is well described by a three component power law. Over the years, it has been argued that the IMF of stellar clusters, at least in the intermediate to high mass end, is 'universal', that is, $\psi_{h}$ is given , within statistical uncertainties, by the Salpeter value and is independent of the environment or of the protocluster cloud properties. Elmegreen (2008) calculated the statistical uncertainties of the IMF slope for randomly sampled IMFs for clusters of different masses. He found that the uncertainties on $\psi_{h}$ are of the order of $\pm 0.15-0.2$ for clusters masses in the range $10^{3}-3\times 10^{4}$ M$_{\odot}$, respectively. However, deviations from 'universality' at both the low and high mass ends have been reported in many observations. At high masses, the IMF is observed to be generally top-heavy in young starburst clusters such as Arches (e.g., Figer et al. 1999; Stolte et al. 2005; Kim et al. 2006), NGC 3603 (e.g., Harayama et al. 2008), and R136 in the central region of 30 Dor (Andersen et al. 2009). Espinoza et al. (2009) re-observed the Arches cluster and obtained an overall slope of $\psi_{h}=-2.1 \pm 0.2$ for the cluster and $\psi_{h}=-1.88 \pm 0.2$ in its inner annulus (within 0.2 pc) whereas Stolte et al. (2005) found a slope of $\psi_{h}= -1.26 \pm 0.07$ in the central region and Stolte et al. (2005) and Kim et al. (2006) obtained slopes that are in the range of $-[1.7-1.9]$ in the second annulus of cluster (depending on whether the entire covered mass range is considered or split between intermediate and high mass stars). Even for less massive clusters, the universality of the IMF is not well established. In fact, determinations of the IMF of many star clusters show cluster-to cluster's variations in $\psi_{h}$ and deviations from the Salpeter value that are in some cases larger than those due to statistical uncertainties (e.g., Massey et al. 1995; Okumura et al. 2000; Massey 2003; Leistra et al. 2005,2006; Sharma et al . 2007). Da Rio et al. (2009) found that the IMF of the stellar association LH 95 in the Large Magellanic Cloud (LMC) has a slope of $\psi_{h} \sim -2$ for stellar masses $M \gtrsim 1.1$ M$_{\odot}$. Many observational groups have attempted to use their determinations of the IMFs of various Galactic and extragalactic clusters in order to favor one or the other of the standard IMFs (i.e., the Chabrier IMF and the Kroupa IMF). For example, Liu et al. (2009)  recently derived the IMF of the NGC 1818 cluster in the LMC and argued that the IMF of this cluster is consistant with both the Chabrier and Kroupa IMFs. A close inspection of the IMF of NGC 1818 (Figures 7 and 8 in their paper) clearly shows the best fit Chabrier IMF to the data would predict a peak at $0.3$ M$_{\odot}$, whereas the data indicates that the peak of the distribution is around $0.8$ M$_{\odot}$. In fact the data shows that there is instead a depression at the predicated peak by the Chabrier IMF at $0.3$ M$_{\odot}$. As for the Kroupa IMF, it systematically overestimates the number of stars  of masses $\gtrsim 1.5$ M$_{\odot}$ in NGC 1818, underestimates the number of stars at the peak, and does not go through the error bars of most of the points. 

Another related issue is that massive stars appear to be preferentially located in the central parts of the clusters (e.g., Hillenbrand \& Hartmann 1998; Figer et al. 1999; Sirianni et al . 2002; Stolte et al. 2002; Gouliermis et al. 2004; Chen et al. 2007, Sharma et al. 2008). The mass segregation of stars in clusters could be due to dynamical two-body relaxation effects (e.g., Vesperini \& Heggie 1997; Kroupa 2001; Mouri \& Taniguchi 2002; McMillan et al. 2007) but also could have a primordial origin. Dib (2007) and Dib et al. (2007a,2008a) showed that the shallower-than Salpeter IMF in the central regions of the Arches cluster, and of likewise starburst clusters, can be reproduced by models based on the rapid coalescence of dense prestellar cores (PSCs) before they collapse and turn into stars. Recently, Chatterjee et al. (2009) confirmed the results of Dib et al. (2007a) as they find that it is unlikely for the mass segregation in the Arches cluster, at its observed current level, and for the cluster age, to be the result of dynamical mass segregation starting from a standard non-segregated Kroupa IMF. They conclude that a shallower IMF in the inner regions of the cluster and which would be the imprint of the star formation process is needed in order to explain the degree of mass segregation presently observed in Arches. 

Undeniably, the origin of the IMF is intimately related to the origin and evolution of the gravitationally bound molecular cloud cores in which stars form. Several observational studies using a variety of techniques and wavelengths have shown that the slope of the dense prestellar core mass function (PSCMF) is well bracketed by the estimated slopes of the IMF (Motte et al. 1998, 2000; Johnstone 2000,2001; Kirk et al. 2006, Stanke et al. 2006; Johnstone \& Bally 2006; Alves et al. 2007). The role played by several physical processes on the origin and evolution of the PSCMF has been extensively studied both theoretically and numerically. The major physical processes considered are: a) gravitational fragmentation (e.g., Zinnecker 1984; Larson 1985; Klessen et al. 1998; Klessen \& Burkert 2001), turbulent fragmentation (Elmegreen 1993; Padoan 1995; Padoan et al. 1997; Padoan \& Nordlund 2002; Padoan et al. 2007), gas accretion (Zinnecker 1982; Larson 1992; Bonnell et al. 1997; Klessen \& Burkert 2000; Bonnell et al. 2001a,b; Basu \& Jones 2004; Bate \& Bonnell 2005; Dobbs et al. 2005; Bonnell \& Bate 2006; Clark \& Bonnell 2006; Banerjee et al. 2006; Bonnell et al. 2007, Heitsch et al. 2008; Clark et al. 2008; Dib et al. 2008b, Offner et al. 2008, Myers 2009), cores or star coalescence (Field \& Saslaw 1965; Nakano 1966; Silk \& Takahashi 1979; Podsiadlowski \& Price 1992; Price \& Podsiadlowski 1995; Bonnell et al. 1998; Bonnell \& Bate 2002; Elmegreen \& Shadmehri 2003; Elmegreen 2004; Shadmehri 2004; Davies et al. 2006; McMillan et al. 2007; Kitsionas \& Whitworth 2007; Dib 2007; Dib et al. 2007a,2008a), ejection of predominantly low mass stars (e.g., Bate et al. 2002; Goodwin et al. 2004; Bate \& Bonnell 2005), and magnetic fields (Shu 2004, Dib et al. 2007b; Nakamura \& Li 2008; Price \& Bate 2008; Kunz \& Mouschovias 2009). Accretion onto PSCs which are on their way to form stars might also be regulated by the effects of feedback from protostellar ouflows and jets (Li \& Nakamura 2006; Nakamura \& Li 2007) and ionization fronts and winds from massive stars (Adams \& Fatuzzo 1996; Dale et al. 2005; Dale \& Bonnell 2008; Kevlahan \& Pudritz 2009). 

\subsection{EVIDENCE FOR ACCRETION IN HIGH MASS STAR FORMING REGIONS ?}\label{observations}

\begin{figure}
\begin{center}
\epsfig{figure=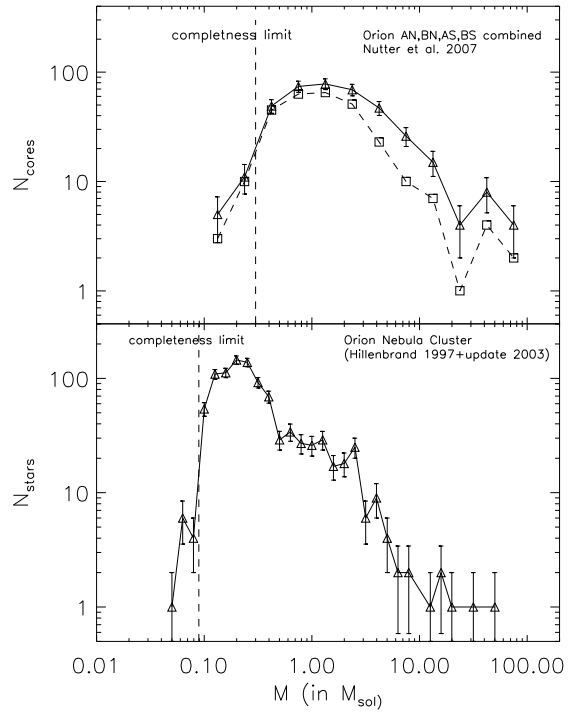,width=\columnwidth}
\end{center}
\vspace{0.5cm}
\caption{Top: Core mass function of the Orion star forming region. The data points combine the cores from the Orion A North and South regions and the Orion B  North and South regions (from Nutter et al. 2007). The full lines include all cores whereas the dashed lines correspond to the population of starless cores (i.e., no young stellar objects detected within them in the infra-red). Bottom: Initial mass function of the Orion Nebula Cluster as plotted by Hillenbrand (1997) and updated by L. Hillenbrand in 2003.}
\label{fig1}
\end{figure}

Among the above mentioned processes, core coalescence and gas accretion might play a central role in modifying the initial PSCMF inherited from the gravo-turbulent fragmentation of a protocluster clump. Dib et al. (2007a) have shown that an efficient coalescence process of dense PSCs in a protocluster clump and their subsequent collapse into stars can result in a significantly shallower-than Salpeter IMF in the intermediate-to-high mass regime. The process based on the coalescence and collapse of dense cores can explain the observed shallow mass functions of young and massive starburst clusters such as Arches, NGC3603 and RC136 (Kim et al. 2006; Harayama et al. 2007;  Andersen et al. 2009; Espinoza et al. 2009). In the case of  less centrally condensed clusters, the coalescence of their precursor PSCs might be less important. However, dense PSCs will continue to accrete gas from their surrounding medium and continue their mass growth. Fig.~\ref{fig1} (top) displays the combined PSCMF of dense cores in Orion A North, Orion A South, Orion B North and Orion B South based on the compilation of cores in Orion by Nutter \& Ward Thompson (2007). Note that the PSCMF displays a tail-like feature at the high-mass end (Another complex PSCMF is the one obtained for for the W3 region as observed by Moore et al. 2007). However, if the high mass cores in the crowded regions OMC1 and NGC 2024 are excluded, the distribution of dense cores in Orion is well fitted with a universal lognormal distribution (Andr\'{e} et al. 2008). Fig.~\ref{fig1} (bottom) displays the IMF of the Orion Nebula Cluster (ONC) as initially derived and plotted by Hillenbrand (1997) and updated by Hillenbrand in 2003. Both the PSCMF of Orion and the IMF of the ONC show an intriguing flattened tail in the high mass regime. Furthermore, The IMF of the ONC displays complex features, among which is a plateau in the mass range 0.45-2.5 M$_{\odot}$, a nearly universal slope between 2.5 and 10 M$_{\odot}$, and another plateau in the high mass regime. Note that a tail at the high mass end is not observed in  the CMF of low mass nearby star forming regions as well as in the IMF of young embedded clusters which do not contain massive stars such as IC 348, NGC 2362 and Chamaeleon I (e.g., Luhman et al. 2000; Luhman et al. 2003a,b; Lada \& Lada 2003; Muench et al. 2003; Luhman 2004,2007).  

Based on the estimate of the lower limit of the ONC mass and the assumption that the IMF of the ONC can be described by a Kroupa IMF, Pflamm-Altenburg \& Kroupa (2006) argued that the ONC should harbor about 40 OB stars that are heavier than 5 M$_{\odot}$ whereas only 10 OB stars are observed to be present in the cluster. Using N-Body simulations, they showed that it would be possible for a compact cluster containing about 40 OB stars to eject 30 of them, along with many lower mass stars, within a timescale of $\sim 1$ Myr. The OB stars should be detected in the mid-infrared by the bow shocks they produce as they evade the cluster. Runaway stars that could have been ejected from dense clusters are observed in star forming regions (e.g., Hoogerwerf et al. 2000; G\'{o}mez et al. 2005). Gvaramadze \& Bomans (2008) report the observations of three bow shocks produced by O-type stars ejected from the NGC 6611 (M16) cluster. Note that Huthoff \& Kaper (2002) discussed the fact that the detection of bow-shocks due to stellar winds from massive stars may depend on the local conditions of the ISM where the star is moving. However, as compelling as the scenario proposed by Pflamm-Altenburg \& Kroupa is, there is no observational evidence that about 30 OB stars have been ejected from the ONC. Even for low mass stars, there are also no observational support for the presence of runaway low-mass stars in the ONC (O'Dell et al. 2005) contrary to what has been suggested by Poveda et al. (2005). 

In this work, we assess how the mass function of PSCs is modified by the effects of time dependent gas accretion. As initial conditions for the distributions of PSCs in a protocluster clump, we use ones resulting from the local turbulent fragmentation of the clump. We show that the PSCMF evolves quickly under the effect of accretion and develops a tail-like feature at the high mass end, comparable to the one observed for the mass function of dense cores in star forming regions which harbor a population of massive cores (i.e., Orion). We also account for the transition from dense cores/protostars to stars and discuss the evolution of both the CMF and IMF as a function of the model parameters. In \S~\ref{protoclusters}, we briefly discuss  the observed properties of star forming protocluster clumps and present our prescription for modeling them. In \S~\ref{core_model}, we discuss some of the properties of the PSCs and in \S~\ref{ini_cond} we present the local mass distributions of PSCs that are formed in the clump uniformly over time. In \S~\ref{acc_turb} we describe our adopted accretion model for accretion onto the cores, and in \S~\ref{co_evol} we present the case of a fiducial model and its application to the Orion Nebula Cluster. Namely, we study the temporal co-evolution of the PSCMF and IMF in the fiducial model and show how the model is able to reproduce simultaneously the complex features in the PSCMF in Orion and the IMF of the ONC. We also address the related issues of age spread of stars and their mass segregation. The effects of varying the model main parameters on the resulting IMFs are presented in \S~\ref{param_study} and in \S~\ref{summary}, we conclude. All the variables of the model are listed and explained in Tab.~\ref{tab1} including the free parameters that are studied in \S~\ref{param_study}.

\section{PROTOCLUSTER CLUMPS}\label{protoclusters}

\subsection{OBSERVATIONS}\label{obs_clusters}

\begin{figure}
\begin{center}
\epsfig{figure=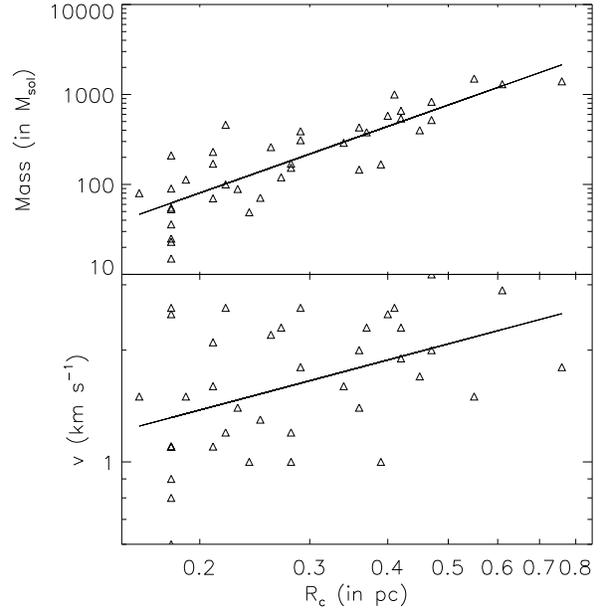,width=\columnwidth}
\end{center}
\vspace{0.5cm}
\caption{Scaling relations for clumps observed in the C$^{18}$O molecular emission line, namely the mass-size relation (top), and the velocity dispersion-size relation (bottom). These figures are based on data presented in figures 4 and 6 in Saito et al. (2007). Over plotted are least-square fits to the data.}
\label{fig2}
\end{figure}

Over the last two decades, several studies using a variety of wavelengths and techniques have established that star clusters form in dense ($\gtrsim 10^{3}$ cm$^{-3}$) clumps embedded in a lower density parental molecular cloud (e.g., Lada \& Lada 2003; Shirley et al. 2003; Minier et al. 2005; Allen et al. 2007 and references therein).  Saito et al. (2007) recently studied, using  the C$^{18}$O molecular emission line, a large sample of cluster forming clumps whose masses and radii vary between [15-1500] M$_{\odot}$Êand [0.14-0.61] pc, respectively. Fig.~\ref{fig2} displays the scaling properties for the star forming clumps observed by Saito et al. (2007) (based on a re-interpretation of figures 4 and 6 in their paper). The mass-size, and velocity dispersion-size relations which we adopt to further constrain our models and which are obtained by performing least square fits to the data points in Fig.~\ref{fig2} (the fits are over-plotted to the data in Fig.~\ref{fig2}), are given by 

\begin{equation}
M_{c}({\rm M_{\odot}})=10^{3.62 \pm 0.14} R_{c}^{2.54 \pm 0.25} ({\rm pc}), 
\label{eq1}
\end{equation}

\noindent and 

\begin{equation}
v_{c}({\rm km~s^{-1}})=10^{0.45 \pm 0.08} R_{c}^{0.44 \pm 0.14}({\rm pc}).
\label{eq2}
\end{equation}

\subsection{MODELS}\label{model_clusters}

In Dib et al. (2007a), we have adopted a protocluster clump model that follows an $r^{-2}$ density profile. Albeit this is a fairly good representation for star forming clumps,  we allow here for generic variations in the clump density profiles by assuming that they can be described by the following function:

\begin{equation} 
\rho_{c}(r)= \frac{\rho_{c0}} {1+(r/R_{c0})^{b}},
\label{eq3}
\end{equation}

\noindent where $R_{c0}$  is the clump's core radius, $\rho_{c0}$ is the density at the center, and $b$ a parameter that accounts for variations in the clump radial density profiles in their outer regions. For a given mass of the clump, the central density is given by the following equation, which can be easily solved numerically:\footnote{An analytical solution to Eq.~\ref{eq4} exists and is given by $\frac {M_{c}}  {4 \pi (1/3) R^{3}_{c}  F^{2}_{1} (3/b,1;1+(3/b);-(R_{c}/R_{c0})^{b}}$, where $F^{2}_{1}$ is a (2,1) order hypergeometric function.}

\begin{equation} 
\rho_{c0}= \frac{M_{c}} { \int_{0}^{R_{c}}  4 \pi r^{2}/ (1+(r/R_{c0})^{b}) dr},
\label{eq4}
\end{equation}

\noindent where $R_{c}$ is the radius of the clump. Note that a variety of profiles have been derived for star forming clumps ranging from $b \sim -1.5$ to $\sim -2.5$ (e.g., Motte et al. 1998). The temperatures of the cluster forming clumps are observed to vary between 15 and 70 K (e.g., Saito et al. 2007). We take a conservative value of the temperature of $T=20$ K, which is probably more representative of their central parts where the bulk of the mass is located (e.g., Minier et al. 2005). In order to further constrain the models and minimize the number of parameters, we relate the size of the protocluster clumps to their mass using the mass-size relation presented in \S~\ref{obs_clusters}. In our models, the proto-cluster clumps are assumed to be in equilibrium. Tan \& McKee (2002) and Tan et al. (2006) argued that cluster-forming clumps can be in a state of equilibrium for a few to several dynamical crossing times. However, as our results below will show, the final IMF of a stellar cluster in a protocluster clump is formed on a timescale which is of the order of $0.2-0.3~t_{ff,c}$ , where $t_{ff,c}$ is the free-fall timescale of the clump. Thus, the equilibrium is required to hold for much shorter timescales tan those suggested by Tan \& McKee (2002). Elmegreen \& Shadmehri (2003) and Shadmehri (2004) assumed that star forming clumps in a molecular cloud are virialized. This might be a plausible hypothesis if the clumps  were indeed the dissipative structures of turbulence in the interstellar medium. However, numerical simulations (e.g., Dib et al. 2007b, Dib \& Kim 2007) show that clumps and cores in molecular clouds are not in virial equilibrium. In the absence of detailed information about the velocity dispersion inside the cores in the Saito et al. (2007) study, we assume that the clump-clump velocity dispersion they derived (i.e., Eq~\ref{eq3}) is also valid on the scale of the clumps themselves and of their substructure. 

\section{THE PRESTELLAR CORES MODEL}\label{core_model}

Whitworth \& Ward-Thompson (2001) have applied a family of Plummer sphere-like models to the contracting prestellar dense core L1554. This core is similar to the population of gravitationally bound cores that can be found in a clump and that are considered in this work. They found a good agreement with the observations of L1554 if the density profile of the core has the following form:

\begin{equation} 
\rho_{p}(r_{p})= \frac{\rho_{p0}}{[1+(r_{p}/R_{p0})^{2}]^{2}},
\label{eq5}
\end{equation}

\noindent where $\rho_{p0}$ and $R_{p0}$ are the central density and core radius of the PSC, respectively. Note that the radius of the PSC, $R_{p}$, depends both on its mass and on its position within the MC. The dependence of $R_{p}$ on $r$ requires that the density at the edges of the PSC equals the ambient clump density, i.e., $\rho_{p}(R_{p})=\rho_{c}(r)$. This would result in smaller radii for PSCs of a given mass when they are located in their inner parts of the cloud. The density contrast between the edge of the PSC and its center is given by: 

\begin{equation} 
{\cal C}(r) = \frac {\rho_{p0}}{\rho_{c} (r)}=\frac {\rho_{p0}} {\rho_{c0}} \left[1+ \left(\frac{r}{R_{c0}}\right)^{b} \right].         
\label{eq6}
\end{equation}

Depending on its position $r$ in the cloud, the radius of the PSC of mass $M$, $R_{p}$, can be calculated as being $R_{p} (r,M)=a(r)~R_{p0} (r,M)$, where: 

\begin{equation} 
R_{p0}(r,M)= \left(\frac{M}{2 \pi \rho_{p0}} \right)^{1/3} \left(\arctan[a(r)]-\frac{a(r)}{1+a(r)^{2}} \right)^{-1/3},
\label{eq7}
\end{equation} 

\noindent and with $a(r)=({\cal C}(r)^{1/2}-1)^{1/2}$. With our set of parameters, the quantity ${\cal C}^{1/2}-1$ is always guaranteed to be positive. The value $R_p(r,M)$ can be considered as being the radius of the PSC at the moment of its formation. The radius of the PSC will decrease as time advances due to gravitational contraction. Both observational (Lee \& Myers 1999; Jessop \& Ward-Thompson 2000; Kirk et al. 2005; Hatchell et al. 2007; Ward-Thompson et al. 2007) and numerical (V\'{a}zquez-Semadeni et al. 2005a; Galv\'{a}n-Madrid et al. 2007; Dib et al. 2008c) estimates of gravitationally bound cores lifetimes tend to show that they are of the order of a few times their free-fall time,  albeit decreasing (but still larger than one free-fall time) when cores are defined with increasingly higher density tracers/thresholds. Thus, we assume that the PSCs contract on a timescale, $t_{cont,p}$ which we take to be a few times their free fall timescale $t_{ff}$, and which is parametrized by: 

\begin{equation} 
t_{cont,p}(r,M)= \nu ~ t_{ff}(r,M)= \nu \left( \frac {3 \pi} {32~G \bar{\rho_{p}} (r,M)} \right)^{1/2},
\label{eq8}
\end{equation}

\noindent where $G$ is the gravitational constant, $\nu$ is a constant $\ge 1$ and $\bar{\rho_{p}}$ is the radially averaged density of the PSC of mass $M$, located at position $r$ in the clump, and which is calculated as being:

\begin{equation} 
\bar{\rho_{p}}(r,M)=\frac{1}{R_{p}(r,M)}~\int^{R_{p}(r,M)}_{0}\frac{\rho_{p0}}{[1+(r_{p}/R_{p0})^{2}]^{2}} dr_{p}.
\label{eq9}
\end{equation}

Thus, the time evolution of the radius of a PSC of mass $M$, located at position $r$ in the cloud is given by a simple contraction law:

\begin{equation} 
R_{p}(r,M,t)=R_p(r,M,0)~e^{-(t/t_{cont,p})}.
\label{eq10}
\end{equation}

One important issue is the choice of the quantity $\rho_{p0}$. In Dib et al. (2007a), we have assumed that the dense PSC have a constant peak density that is independent of the PSC mass and position. In this work, we adopt a slightly more realistic approach along the following lines: We first assume that the minimum density contrast that should exist between the center of the PSC and its edge is of the order of the critical Bonnor-Ebert sphere value and that is $\gtrsim 15$. Secondly, we assume that the density contrast between the center and the edge of the PSCs depends of their masses following a relation of the type:

\begin{equation}
\rho_{p0}  \propto M^{\mu}.
\label{eq11}
\end{equation}

\begin{figure}
\begin{center}
\epsfig{figure=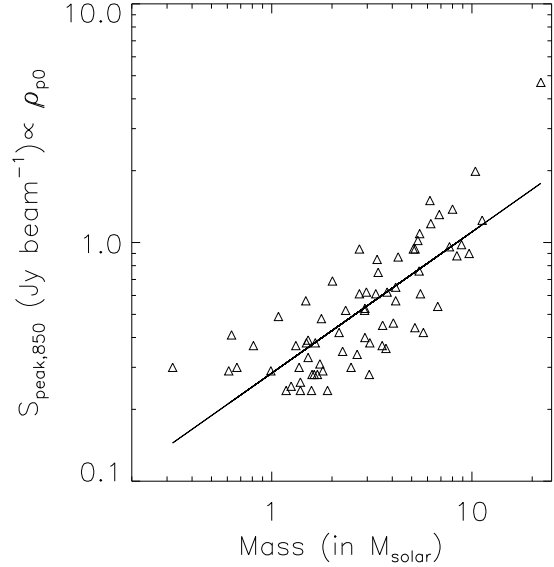,width=\columnwidth}
\end{center}
\caption{The variation of the peak emission in the 850$\micron$ and hence of the peak density (in our paper named $\rho_{p0}$) as a function of the mass for a sample of dense cores in the observations of Johnstone \& Bally (2006). A least square fit to the data points yields $S_{850,peak} \propto \rho_{p0} \propto M^{0.59\pm 0.05}$.}
\label{fig3}
\end{figure}

\noindent Caselli \& Myers (1995) found that massive PSCs in the star forming regions L1641 (Orion A) and L1630 (Orion B)  are denser than lower mass ones. The data points in Fig.~\ref{fig3} display the peak dust continuum emission at $850 \micron$, $S_{850,peak}$ of dense cores as a function of their masses  as observed by Johnstone \& Bally (2006) in the Orion B molecular cloud. The dust continuum emission is directly proportional to the dust density, and thus, to the gas density, assuming the gas-to-dust ratio is constant and independent of the density. A fit to the observed data points in Fig.~\ref{fig3} yields the following relation:

\begin{equation}
S_{850,peak}\propto \rho_{p0}\propto M^{\mu=0.59\pm 0.05}.
\label{eq12}
\end{equation}  

\noindent Thus, the density contrast between the center and the edge for a PSC with the minimum mass we are considering, $M_{min}$  (typically $M_{min}=0.1$ M$_{\odot}$) is 15, whereas for a more massive PSC of mass $M$, the density contrast will be equal to $15\times (M/M_{min,})^{\mu}$, with $\mu$ assumed to be in the range $[0-0.6]$ .  
 
\section{THE MASS DISTRIBUTION OF GRAVITATIONALLY BOUND PRESTELLAR CORES} \label{ini_cond}

As initial conditions for the PSCs mass distribution at different clump radii, we adopt distributions that are the result of the gravo-turbulent fragmentation of the clump. As in our previous work (Dib et al. 2007a), we use the formulation given by Padoan \& Nordlund (2002, PN02) in order to calculate the local distributions of PSCs masses. For standard parameters characteristic of star forming regions, the PN02 formalism gives a PSCMF for PSCs that bears a good resemblance to the observations. Hennebelle \& Chabrier (2008) proposed a derivation of the PSCMF based on the Press-Schechter formalism applied to dense cores. Yet,  the differences between the two models as far as gravitationally bound cores are concerned, are quite marginal. In the following, we briefly remind what the ingredients of the PN02 model are. The model assumes that the probability distribution function of an isothermal, turbulent, compressible gas is well described by a lognormal distribution (V\'{a}zquez-Semadeni 1994) and is given by:

\begin{equation}
P(ln~x) d~\ln~x=\frac{1}{\sqrt{2 \pi \sigma_{d}}} \exp \left[-\frac{1}{2} \left(\frac{\ln x-\bar{\ln x}}{\sigma_{d}} \right)^{2} \right] d~\ln~x,
\label{eq13}
\end{equation}

\noindent where $x$ is the number density normalized by the average number density, $x=n/\bar{n}$. The standard deviation of the density distribution $\sigma_{d}$ and the mean value $\bar {\ln x}$ are functions of the local thermal rms Mach number, $\cal M$ and  $\bar{\ln x}=-\sigma^{2}_{d}/2$ and $\sigma^{2}_{d}=\ln(1+{\cal M}^{2} \gamma^{2})$. PN02 suggested a value of $\gamma \sim 0.5$, whereas Kritsuk et al. (2007) using higher resolution simulation found that $\gamma \sim 0.260 \pm 0.001$. The latter value is the one adopted in our models. A second step in this approach is to determine the mass distribution of dense cores. PN02 showed that by making the following assumptions: (a) the power spectrum of turbulence is a power law and, (b) the typical size of a dense core scales as the thickness of the post-shock gas layer, the cores mass spectrum is given by:

\begin{equation}
N(M)~d~log~M \propto M^{-3/(4-\beta)} d~\log~M,
\label{eq14}
\end{equation}   

\noindent where $\beta$ is the exponent of the kinetic energy power spectrum, $E_{k} \propto k^{-\beta}$, and is related to the exponent $\alpha$ of the size-velocity dispersion relation in the cloud with $\beta=2 \alpha+1$. However, Eq.~\ref{eq14} can not be directly used to estimate the number of cores that are prone to star formation. It must be multiplied by the local distribution of Jeans masses. At constant temperature, this distribution can be written as:

\begin{equation}
P(M_{J})~dM_{J}=\frac{2~M_{J0}^{2}}{\sqrt{2 \pi \sigma^{2}_{d}}} M^{-3}_{J} \exp \left[-\frac{1}{2} \left(\frac{ln~M_{J}-A}{\sigma_{d}} \right)^{2} \right] dM_{J},
\label{eq15}
\end{equation}   

\noindent where $M_{J0}$ is the Jeans mass at the mean density $\bar{n}$. Thus, locally, the number of cores is given by: 

\begin{eqnarray}
N (r,M)~d\log~M =f_{0}(r)~M^{-3/(4-\beta)} \nonumber \\
            \times \left[\int^{M}_{0} P(M_{J}) dM_{J}\right]d\log~M,
\label{eq16}
\end{eqnarray}   

Eq.~\ref{eq16} can be solved analytically yielding the following form:

\begin{equation}
N(r,M)~dm=f_{0}(r)~\left[1+erf\left( \frac{4 \ln M+\sigma_{d}^{2}}{2 \sqrt{2} \sigma_{d}}\right) \right]  M^{3/(4-\beta)} dm,
\label{16bis}
\end{equation}

where 'erf' is the error function. The local normalization coefficient $f_{0}(r)$ is obtained by requiring that $\int^{M_{max}}_{M_{min}} N (r,M)~dM=1$ in a shell of width $dr$, located at distance $r$ from the clump's center. Therefore, the local distribution of cores generated in the clump, at an epoch $\tau$, $N(r,M,\tau)$, is obtained by multiplying the local normalized function $N(r,M)$ by the local rate of fragmentation such that:

\begin{equation}
{N} (r,M,\tau)~dt=\frac{\epsilon_{c}(r) \rho_{c}(r)} {<M>(r)~t_{cont,p} (r,M)} \frac{dt}{t_{ff,cl}} N(r,M),
\label{eq17}
\end{equation} 

\noindent where $dt$ is the time interval between two consecutive epochs, $<M>$ is the average core mass in the local distribution and is calculated by $<M>=\int_{M_{min}}^{M_{max}} M~N (r,M,0)~dM$, and $\epsilon_{c}$ is a parameter smaller than unity which describes the local mass fraction of gas that is transformed into PSCs per free fall time of the protocluster clump, $t_{ff,cl}$ (i.e., core formation efficiency per free fall time).  In principle, $\epsilon_{c}$ might have a radial dependence, but for the sake of simplicity, we shall assume $\epsilon_{c}$ to be a constant, independent of radius. In our model, since PSCs are generated uniformly over time in the clump, $\epsilon_{c}$ is also independent of time.  
 
\begin{figure}
\begin{center}
\epsfig{figure=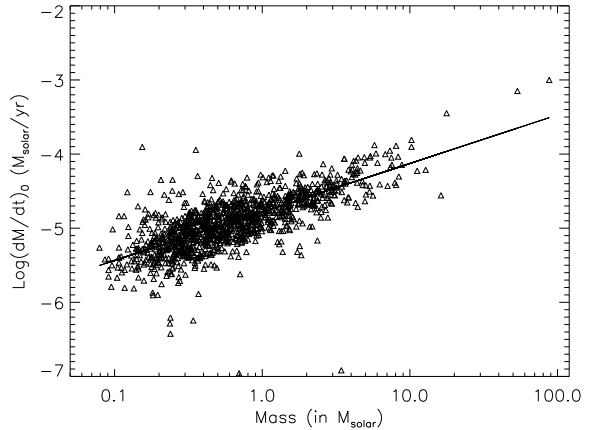,width=\columnwidth}
\end{center}
\caption{The relationship between the core mass and the accretion rate normalization coefficient. This figure is based on the combined data of Figure 3 in Schmeja \& Klessen (2004). $\dot{M}_{0}$ is related to the maximum temporal accretion rate by $log\dot{M}_{0}=log\dot{M}_{max}$+7. Overplotted to the data is a linear fit whose parameters are given in the text.}
\label{fig4}
\end{figure}
 
\section{THE ACCRETION MODEL: ACCRETION IN A TURBULENT MEDIUM}\label{acc_turb}

Self-gravitating pretstellar cores can accrete mass from their surrounding environment (e.g., Klessen 2001; Padoan et al. 2005; Schmeja \& Klessen 2004, Dib et al. 2008b). As they collapse, they also produce bipolar outflows and jets which will tend to reduce the accretion rate by removing a fraction of the available mass surrounding them (e.g., Tomisaka 2002). Thus, the net effect of gas accretion by the core and of gas dispersal by its bipolar outflow is given by $\dot{M}=\dot{M}_{acc}-\dot{M}_{loss}$, where $\dot{M}_{acc}$ and $\dot{M}_{loss}$ are the mass accretion and mass loss rates, respectively. Accretion rates onto PSCs are difficult to measure directly from the observations. Observations of the accretion rates are mostly obtained in the protostellar phase and even then, they are usually estimated indirectly from the spectral energy distribution or eventually from the properties of associated outflows (i.e., a correlation between the accretion rate and outflow strength, Hartignan et al. 1995; Bontemps et al. 1996; Wolf-Chase et al. 2003). The accretion rates are observed to vary as the PSC core evolves. Typical accretion rates for Class 0 protostars are found to be in the range $10^{-5} \lesssim \dot{M}_{acc}/$ M$_{\odot}$ yr$^{-1}$ $\lesssim 10^{-4}$ (Hartmann 1998; Narayanan et al. 1998; Andr\'{e} et al. 1999; Ceccarelli et al. 2000; Jayawardhana et al. 2001; Di Francesco et al. 2001; Maret et al. 2002; Beuther et al. 2002a,b), whereas accretion rates of Class I protostars are typically an order of magnitude smaller (Henriksen et al. 1997; Andr\'{e} et al. 2000) with values ranging between $\sim 10^{-7}$ M$_{\odot}$ yr$^{-1}$ and $5\times 10^{-6}$ M$_{\odot}$ yr$^{-1}$ (e.g., Brown \& Chandler 1999; Greene \& Lada 2002; Boogert et al. 2002; Young et al. 2003). Massive cores that are likely to form O stars are observed to have even higher accretion rates in the range of $10^{-3}-10^{-2}$ M$_{\odot}$ yr$^{-1}$ (Ho \& Young 1996; Zhang \& Ho 1997; Sandell et al. 2005; Beltr\'{a}n et al. 2006; Garay et al. 2007; Zapata et al. 2008). 

In the prestellar core phase, Padoan et al. (2005) measured the accretion rate of individual cores in numerical simulations of turbulent and self-gravitating molecular clouds. They argued that the observed accretion rates of prestellar cores and protostars can be explained by a volume averaged Bondi-Hoyle accretion rate $\dot{M}_{BH}$ (Bondi \& Hoyle 1944). However, it remains unclear whether the Bondi-Hoyle formalism, which describes the accretion onto a point mass from a homogeneous gas distribution and with no (or a uniform) velocity field, is suited for the case of gas accretion by PSCs in a turbulent molecular clump. In a protocluster clump, the gas surrounding the PSCs is highly inhomogeneous, and the velocity field could have a rather complex topology (e.g., Dib et al. 2007b; V\'{a}zquez-Semadeni et al. 2008). Krumholz et al. (2006) showed that for an ensemble of accreting objects from a turbulent medium, some of the objects will accrete in a fashion that is closer to the Bondi-Hoyle mode if the gas surrounding them has a well ordered velocity field and a low vorticity level, whereas other objects will accrete according to an accretion rate in the vorticity dominated regime with an accretion rate $\dot{M}_{w}$  (derived by Krumholz et al. 2005, and approximated by Krumholz et al. 2006). Krumholz et al. (2006) suggested that the Bondi-Hoyle accretion rate be replaced by $\dot{M}_{turb} \sim [\dot{M}_{BH}({\bf x})^{-2}+\dot{M}_{w}({\bf x})^{-2} ]^{-1/2}$. However, the accretion rate formula proposed by Krumholz et al. (2006) does not take into account the self-gravity of the gas. Klessen (2001) derived time dependent accretion rates onto cores embedded in a self-gravitating gas and proposed to describe them by empirical fit functions with a dichotomy of the fit parameters covering four distinct mass ranges. Schmeja \& Klessen (2004, SK04) measured the time dependent accretion rates onto cores in their simulations of turbulent and self-gravitating molecular clouds. Using data from a large ensemble of cores (i.e., $\sim 6000$ cores), they proposed an empirical fit to their measurements of the accretion rates which has the following functional form:  

\begin{equation} 
\log \dot{M}_{SK04}(M,t)=\log \dot{M_{0}}~\frac{e}{\tau_{acc}}~t~e^{-t/\tau_{acc}},
\label{eq18}
\end{equation}

\noindent where $\tau_{acc}$ is a parameter which describes the timescale over which the accretion rate declines from its maximum value and $\dot{M}_{0}$ is a parameter which is linked to the mass of the accreting PSC at the end of the accretion process. SK04 found $\tau_{acc}$ to be smaller than the PSCs free fall time $t_{ff}$, and by averaging over the populations of cores in different mass ranges, they obtained a value of $\tau_{acc} \sim t_{ff}/3$. The second parameter, $\dot{M}_{0}$, is related to the maximum value of the accretion rate by $\log \dot{M}^{max}_{fit}=\log \dot{M_{0}}-7$. SK04 did not provide a relationship between $\dot{M}^{max}_{fit} $ and the temporal mass of the cores. Instead, in figure 3 of their paper, SK04 plotted the values of $\dot{M}^{max}_{fit}$ as a function of the mass of the core at the end of accretion process. However, that for many cores what SK04 call $M_{end}$ is not necessarily the final mass of the cores as they can be in a phase of active accretion when the simulation was terminated. The data in their figure 3 corresponds to a sample of molecular cloud simulations in which turbulence was driven with a variety of Mach numbers (i.e., between 0.1 and 10) and of driving length scales (between half and an eighth of the box size). In Fig.~\ref{fig4} we combined all the data points of figure 3 of SK04 and fitted the $\dot{M}_{max}-M_{end}$ relation. We found :  

\begin{equation}
\dot{M}_{max} = 10^{-4.78\pm0.008} M_{end}^{0.65\pm 0.17}  \rm {M_{\odot}yr^{-1}}.
\label{eq19}
\end{equation}

By adopting the time dependent accretion rate formula of SK04\footnote{Schmeja \& Klessen (2004) describe the collapsing cores (which are replaced by a sink particle) in their simulations as protostellar whereas we call our cores at the moment they are formed in the clump as prestellar. Note however, that the prestellar cores we consider in this work are contracting very rapidly (Eq.\ref{eq10}) and will go through a protostellar phase before forming stars. The cores both in our work and in SK04 are not resolved, and thus, the accretions rates they derive for accretion onto the sink particles in their simulations can be safely applied to describe accretion onto a prestellar cores or onto a protostellar cores (protostar+envelope).  Whether they are termed prestellar cores, protostellar cores, or simply cores is not very relevant. The underlying physics is the same in the sense that these are point-like objects  that are accreting gas in a turbulent and self-gravitating medium that can is either available in the immediate neighborhoud or that is accreted from further away. Would the sink particle in SK04 be resolved such as to have a real protostar, a disk, and an envelope, one could then consider the accretion separately on the protostar or on the protostar+envelope, bearing in mind that these two quantities might simply be directly connected. This is however not the case, and the distinction need not to be made.} (i.e., Eq~\ref{eq18}), we make the approximation that for an accreting core, the actual mass $M(t)$ at each epoch is a new final mass which requires a new normalization of the maximum accretion rate and thus of $\dot{M}_{0}$. Finally, note that the accretion rates in Schmeja \& Klessen (2004) are relative to an average density of $10^{5}$ cm$^{-3}$ which is the average number density adopted in their simulations. Thus, in order to calculate the accretion rates of PSCs located at different positions in a protocluster clump, it is necessary to take into account the effect of the varying background density. Since accretion is directly proportional to the external density, it is necessary to scale the accretion rate of a PSC of a given mass located at a position $r$ in the cloud where the local number density is $n(r)$ by the value of the accretion rate at the $10^{5}$ cm$^{-3}$ number density such that:

\begin{equation} 
\dot{M}_{acc}(r=r(n,M,t))=\frac{n ({\rm cm^{-3}})} {10^{5} (\rm cm^{-3})} \dot{M}_{SK04}(M,t).
\label{eq20}
\end{equation}

\noindent Finally, PSCs also re-inject into the cloud or the larger ISM a fraction of their accreted mass in the form of outflows and jets (Nakano et al. 1995; Matzner \& McKee 2000; Tomisaka 2002; Machida et al. 2007) . Thus, it is also important to include the mass loss by outflows . We moderate the accretion rates by introducing a mass loss rate due to outflows rate such that $\dot{M}_{loss}=0.1~\dot{M}_{acc}$, which leads to an effective accretion rate $\dot{M}_{acc,eff}=0.9~\dot{M}_{acc}$, where $\dot{M}_{acc}$ is given by Eq.~\ref{eq20}.  \\

\section{THE CO-EVOLUTION OF THE PRESTELLAR CORE MASS FUNCTION AND THE IMF}\label{co_evol}

\begin{figure}
\begin{center}
\epsfig{figure=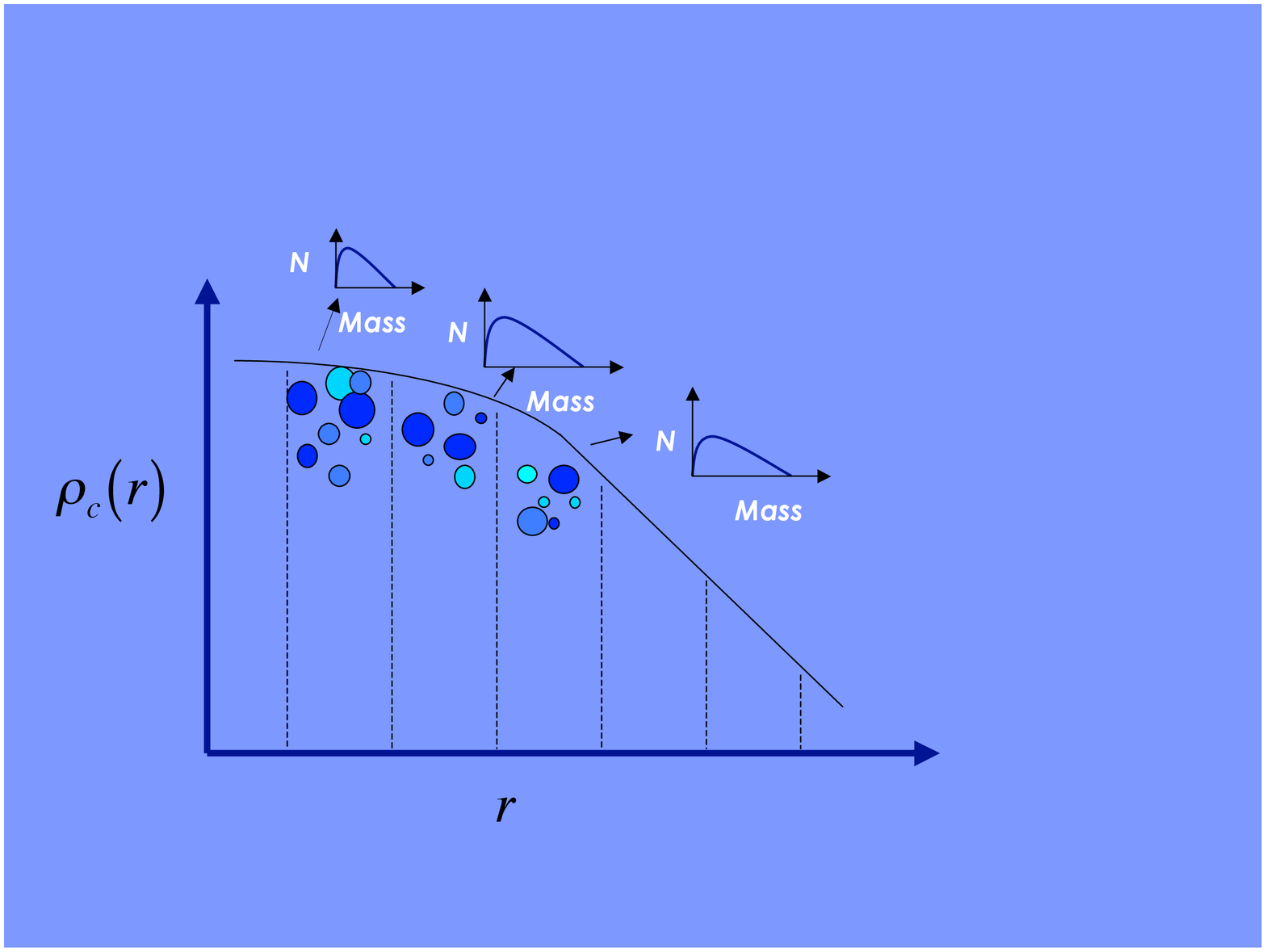,width=0.8 \columnwidth, angle=90}
\end{center}
\caption{A schematic figure showing the set up used in this paper. The circles of different colors represent cores of different ages (i.e., injected at different epochs) along with their local cumulative mass function. The mass function of cores in the entire proto-cluster clump is the sum of all the local distributions.}
\label{fig5}
\end{figure}

\begin{figure*}
\begin{center}
\includegraphics[height=18.5cm, width=15.5cm]{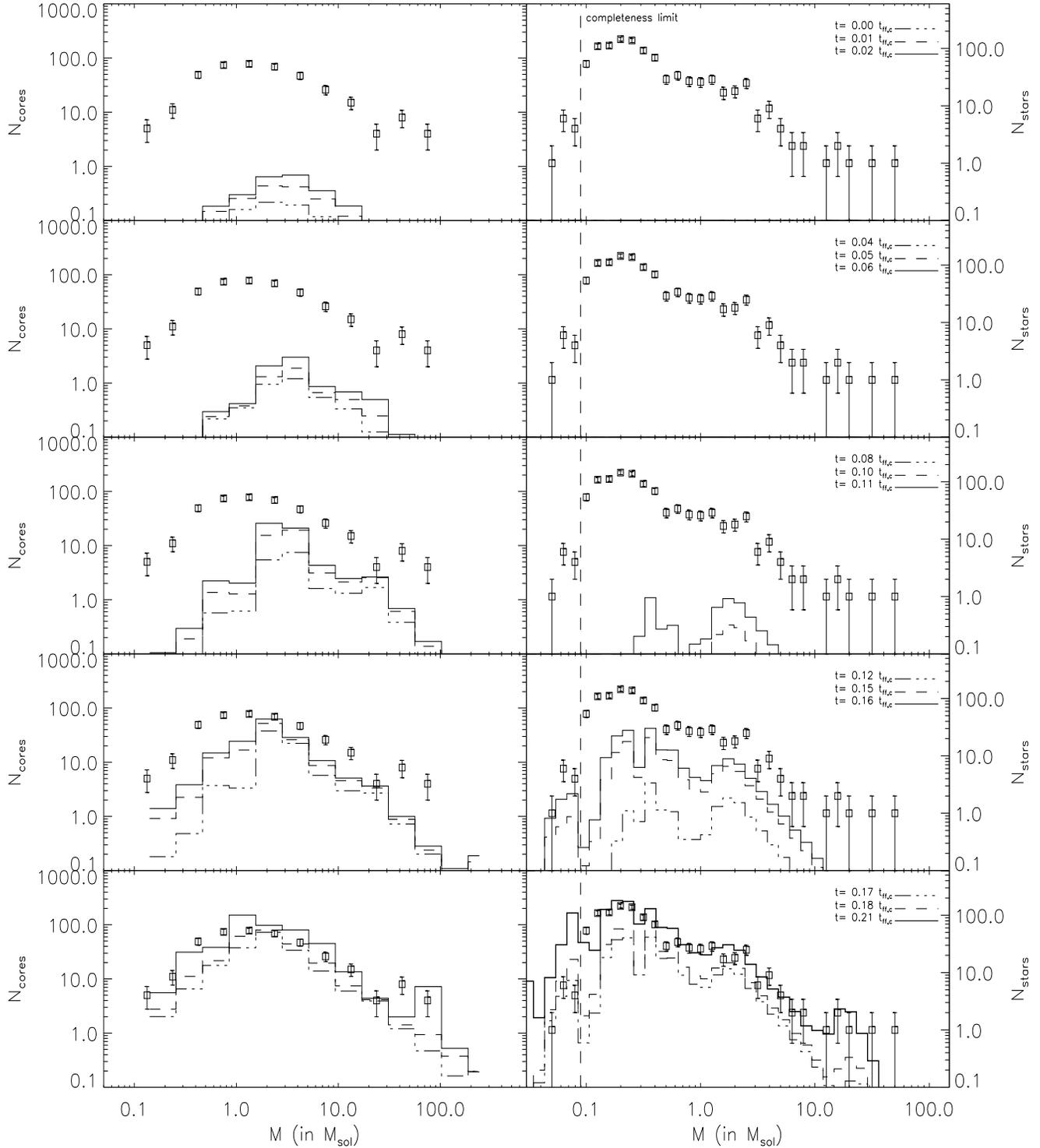}
\end{center}
\vspace{0.7cm}
\caption{ Time evolution of the pre-stellar core mass function (left), and stellar mass function (right) in the protocluster clump with the fiducial model parameters. The stellar mass function is compared to that of the Orion Nebula Cluster (ONC, Hillenbrand 1997; updated by Hillenbrand 2003). The combined dense core mass function of the four regions in Orion which contain dense cores is also shown (left column). The best fit to the data of the ONC is reproduced at $t \sim 0.21~t_{ff,c}$, where $t_{ff,c}$ is the free-fall time of the protocluster clump and which corresponds to the epoch at which gas is expelled from the protocluster clump (thick line in lower right panel). The size of  the mass bin in the model has been re-adjusted such as to resemble the observations mass bins. The PSCMF of the model has been scaled up by a factor of 15 to account for the difference between the clump mass considered in the model ($10^{4}$ M$_{\odot}$) and the combined mass of the regions A and B in Orion which is $\sim 1.5-2 \times 10^{5}$ M$_{\odot}$ (e.g. Nutter \& Ward-Thompson (2007).}
\label{fig6}
\end{figure*}

If the accretion rate $\dot{M}_{acc,eff}$ is generally a function of time and of the mass of the PSC, it can be easily shown that the time variation of a population of PSCs of mass $M$, located at a distance $r$ from the center of the clump, and that was injected into the protocluster clump at the epoch $\tau$, is given, at time $t$, by:

\begin{eqnarray} 
\left(\frac{dN(r,M,\tau,t)}{dt}\right)=\nonumber \\
\left[-\left(\frac{\partial N}{\partial M} \right) \dot{M}_{acc,eff}-\left(\frac{\partial \dot{M}_{acc,eff}}{\partial M}\right) N\right] (r,M,\tau,t).
\label{eq21}
\end{eqnarray}

Whenever a population of PSCs of a given mass $M$, located at a distance $r$ from the center of the cloud has evolved (and accreted) for a time that is equal to its contraction timescale, this population of PSCs is collapsed into stars and the accretion process onto these new born stars is terminated.  In the present model, we do not take into account any further potential sub-fragmentation of the collapsing core. Thus, a single core leads to the formation of a single star/star system and as a consequence the derived IMFs can be compared to system IMFs rather than to single star IMFs. This approach is similar to the one adopted in Dib et al. (2007a) for the study of the effects of cores coalescence on the mass spectrum in which the criteria to turn a PSC into a star was a comparison of the contraction timescale to the time dependent coalescence timescale. The difference between the approach adopted in Dib et al. (2007a) and the one adopted in the present model is the following: In Dib et al. (2007a), the local fraction of mass of the protocluster clump that was injected into PSCs was a quantity that was fixed at the beginning of the model and there was no new generations of PSCs that were injected beyond the initial timestep. In the present model, PSCs with a given mass spectrum dictated by the local dynamical conditions are uniformly injected over time according to the specified rate of PSCs formation per unit free-fall time of the protocluster clump. Thus, the local number of PSCs of a given mass, at a given epoch, is the sum of all the local populations of PSCs of the same mass that have been injected at all epochs that are anterior or equal to the considered epoch (a schematic figure of the clump and the populations of cores is shown in Fig.~\ref{fig5}). Note that the local populations of PSCs of various ages are evolved separately as they are each in a different phase of their accretion history, and that they will collapse and form stars at various epochs. Thus, Eq.~\ref{eq21} is simultaneously solved for all the populations of PSCs whose ages are older or equal to the current epoch. The total local number of PSCs of a given mass $M$, at a time $t$, will be given by:

\begin{equation} 
N(r,M,t)=\sum_{\tau_{i}  \leq t} N(r,M,\tau_{i},t).
\label{eq22}
\end{equation}

As in Dib et al. (2007a), we also assume that only a fraction of the mass of a PSC ends up locked in the star. This implies that a fraction of the mass of the core is re-dispersed into the protocluster clump. In addition to the effect of the protostellar outflows, the rest of the mass is lost  when the stars are formed by the effects of radiation pressure and stellar winds. We account for this mass loss in a purely phenomenological way by assuming that the mass of a star which is formed out of a PSC of mass $M$ is given by M$_{\star}=\xi M$, where $\xi \le 1$. Matzner \& McKee (2000) showed that $\xi$ can vary between $0.25-0.7$ for stars in the mass range $0.5-2$ $M_{\odot}$. It is unknown whether this result holds at higher masses. However, the similarity between the IMF and the dense cores mass function observed by Alves et al. (2007) in the Pipe Nebula might be an indication of a constant $\xi$ across the mass spectrum (i.e., in their case it is $\xi \sim 1/3$; albeit it should be noted that the cores in the Pipe nebula are not believed to be all gravitationally bound). In the absence of strong observational and theoretical constraints, we shall assume that $\xi$ is independent of the mass. We assume that the accretion process onto all PSCs in the protocluster clump is terminated whenever the kinetic energy exceeds the gravitational energy of the clump, resulting in the gas being expelled from the protocluster or at least that star formation is severely decelerated. Similarly to Dale \& Bonnell (2008), we consider that only stars whose mass $M_{\star}$ exceeds $10$ M$_{\odot}$ loose mass with a mass loss rate, $\dot{M}_{\star}$, given by:

\begin{equation}
\dot{M}_{\star}=10^{-5} \left( \frac{M_{\star}}{30~\rm{M_{\odot}}} \right)^{4} \rm{M_{\odot}}~\rm{yr^{-1}}.
\label{eq23}
\end{equation}

We also consider that the terminal velocity of the wind is given by $v_{\inf}=10^{3}$ km s$^{-1}$. The kinetic energy from winds is thus calculated as being

\begin{equation}
E_{wind} = \int_{t'=0}^{t'=t} \int_{m=10\rm{M_{\odot}}}^{m=120~\rm{M_{\odot}}} \left( \frac{N(m) \dot{M_{\star}} (m) v_{\inf}^{2}}{2} dm\right) dt'. 
\label{eq24}
\end{equation}

We assume that only a fraction of $E_{wind}$ will be transformed into systemic motions that will oppose gravity and participate in the evacuation of the bulk of the gas from the proto-cluster clump. The rest of the energy is assumed to be dissipated or carried away from the protocluster clump by a small fraction of the mass, particularly if the massive stars are not born in the center of the clump (see confirmation of this in \S~\ref{mass_segreg}). Thus the relevant energy, $E_{k,wind}$, is given by:

\begin{equation}
 E_{k,wind}=\kappa~E_{wind},
 \label{eq25}
 \end{equation}
 
\noindent where $\kappa$ is a quantity $\leq 1$. It is very difficult to estimate $\kappa$ as its exact value will vary from system to system depending on the number of massive stars, their locations, and the interactions of their winds. It is a quantity that can only be determined by numerical simulations. As a conservative guess for the fiducial model, we take $\kappa=0.1$. $E_{k,wind}$ is compared at every timestep to the absolute value of the gravitational energy, $E_{g}$, which is calculated as being:

\begin{equation}
E_{grav} =  -\frac{16}{3} \pi^{2} G \int_{0}^{R_{c}} \rho_{c}(r)^{2} r^{4}  dr,
\label{eq26}
\end{equation}

\noindent where $\rho_{c}$ is given by Eq.~\ref{eq3}. Note that in this work we only take into account feedback from massive stars in the form of stellar winds. Another important source of feedback from massive stars is their ionizing radiation (e.g., Whitworth 1979, Dale et al. 2005, Lee \& Chen 2007; Gritschneder et al. 2009, Bisbas et al. 2009) which we intend to include in a future work. Finally, we account for the possible modification of the stellar mass function by the effect of stellar winds at the high mass end. The variations in the IMF at the high mass end will be given by:

\begin{eqnarray} 
\left(\frac{dN_{\star}(r,M,t)}{dt}\right)= \nonumber \\
\left[\left(\frac{\partial N_{\star}}{\partial M_{\star}} \right) \dot{M_{\star}}+\left(\frac{\partial \dot{M_{\star}}}{\partial M_{\star}}\right) N_{\star}\right] (r,M,t),
\label{eq27}
\end{eqnarray}

\noindent where $\dot{M}_{\star}$ is the stellar mass loss rate given by Eq.~\ref{eq18}, and $N_{\star}$(r,M,t) is the local number of stars, at time $t$, of mass $M_{\star}$. The model variables and free parameters are summarized in Tab.\ref{tab1}. 

\subsection{A FIDUCIAL MODEL AND COMPARISON TO THE ORION NEBULA CLUSTER}\label{fiducial}

As stated in \S~\ref{observations}, the comparison of our models to the Orion star formation region is motivated by the fact that Orion is the only nearby star forming region which is harboring massive cores, in addition to be relatively well sampled in the regime of low mass cores. The other nearby star forming regions (e.g., Ophiucus, Perseus, Taurus, Pipe Nebula) and for which a PSCMF has been determined, are not known to host massive cores that could be the progenitors of massive stars. Most importantly, on the stellar side, the IMF of the ONC has been obtained by a detailed spectroscopic survey (Hillenbrand 1997; Hillenbrand \& Carpenter 2000) and thus, the features of the IMF of the ONC are much more reliable then IMF determinations based on photometric surveys. The IMF of the ONC also sample a mass range that extends from The data of the PSCMF of Orion displayed in Fig.~\ref{fig1} (top) is the result of the combination of several sub regions of Orion where dense cores are present (Orion A North, Orion A South,  Orion B North, Orion B South) and the local dynamical conditions might be different from one region to another. Thus any theoretical model of the co-evolution of PSCMF and of the IMF in Orion should focus on reproducing, as its primary goal, the IMF of the ONC only as it is the best constrained observable quantity. 

In this section, we discuss the time evolution of the PSCMF and the transition to the IMF for a fiducial model. The parameters for this model are: the mass of the protocluster clump is $M_{c}= 10^{4}$ $M_{\odot}$,  the temperature of the gas is assumed to be $T=20$ K, the clump's core radius and radial density profile exponent are $R_{c0}=0.2$ pc and  $b=-2$, respectively, the exponent of the peak density-mass relation is $\mu=0.2$, the mass fraction of the clump mass that is transformed into PSCs per free-fall time is $\epsilon_{c}=0.07$, the exponent of the velocity dispersion-size relation is $\alpha=0.44$ (from Saito et al. 2007), the ratio of the contraction timescale to the free-fall time of the cores is $\nu=1.8$. and the fraction of the mass of the cores that ends up locked into stars after they collapse is $\xi=0.1$. A value of $\xi=0.1$ is motivated by the fact that the peak of the PSCMF function in Orion is $\sim 2$ M$_{\odot}$, whereas the peak of the IMF of the ONC is located at $\sim 0.2$ M$_{\odot}$. With a mass of $10^{4}$ M$_{\odot}$, the clump has a radius $R_{c}=1.42$ pc according to Eq.~\ref{eq1}. 

We solve the model's equations using a finite differences scheme on a (180,180) linear and logarithmic grid in radius and mass, respectively. The timestep is chosen to be $dt=t_{ff,c}/300$, where $t_{ff,c}$ is the initial free-fall time of the protocluster clump\footnote{We have checked, for our adopted timestep size, that a grid size of $\gtrsim 120$ cells in mass is needed to ensure the convergence of the results}. At any given epoch $t=i \times dt$ (where $i$ is an integer $\leq 300$), we solve $i$ times Eq.~\ref{eq22}. Fig.~\ref{fig6} displays the time evolution of the PSCs populations in the entire protocluster clump (left column) in the fiducial model. Once PSCs of a given  mass $M$, of a given age $\tau$, located at a given distance $r$ from the center of the clump collapse to form stars, they are transferred into the IMF (right column). For comparison, we also over-plot to the model the PSCMF of the four combined regions of Orion (left column) and the IMF of the ONC which we aim to reproduce (right column). 

The overall effect of accretion is to shift  the characteristic mass to higher masses and also to create more massive cores (and as a consequence more massive stars) then were present if the core mass function was a simple pile up of cores formed at different epochs with no accretion involved. However, as cores have finite lifetimes, they will be turned into stars as time goes by, thus emptying the corresponding bins in the PSCMF. For a given population of cores born at the same time, and depending on the value of $\mu$, more massive cores will collapse faster than their lower mass counterparts (for $\mu > 0$). This will lead to a concentration 

In the early phases, the effects of accretion are not visible on the PSCs mass function. The only effects at this stage are a pile-up of the consecutives PSCs populations that are being injected uniformly as time advances. When $t \sim 0.04-0.06~t_{ff,c}$, the effects of accretion become visible with the widening of the plateau around the peak value and the generation of a larger fraction of more massive cores. By $t \sim 0.08-0.11~t_{ff,c}$, the plateau has advanced until masses of $M\sim 20-25$ M$_{\odot}$. A peak in the PSCs distribution is also formed at $t \sim 0.12-0.15~t_{ff,c}$ at $M \sim 0.5$ M$_{\odot}$. At $t\sim 0.08~t_{ff,c}$, the first generations of PSCs collapse into stars and the IMF becomes populated (right column). Since in this fiducial model, the more massive PSCs are more centrally condensed than their less massive counterparts (i.e., $\mu=0.2 > 0$), this leads to the formation of the intermediate mass stars first from the first generations of cores before low mass stars are formed (there are at this stage no massive cores to form massive stars yet). Accretion continues to affect the subsequent generations of intermediate mass cores leading to the formation of massive cores and to the flattening of the PSCs mass function at the high mass end. As time advances, the PSCs mass function continues to flatten at the high mass end and develops a tail-like structure for PSCs masses $\gtrsim 50$ M$_{\odot}$. In the lowest left quadrant, the PSCMF of Orion has been scaled down by a factor of 15 which is roughly the scaling factor between the mass of the protocluster clump we consider here ($10^{4}$ M$_{\odot}$) and the combined masses of the four Orion regions ($1-2 \times10^{5}$ M$_{\odot}$, e.g., Hillenbrand 1997). Although, we have argued earlier that an exact match between the combined PSCMF of the four Orion regions and of the models is not to be expected because of variations that may exist in the local distributions in each of those Orion regions, the agreement between the observations and the models is surprisingly very good at $t \sim 0.21~t_{ff,c}$. 

\begin{figure}
\begin{center}
\epsfig{figure=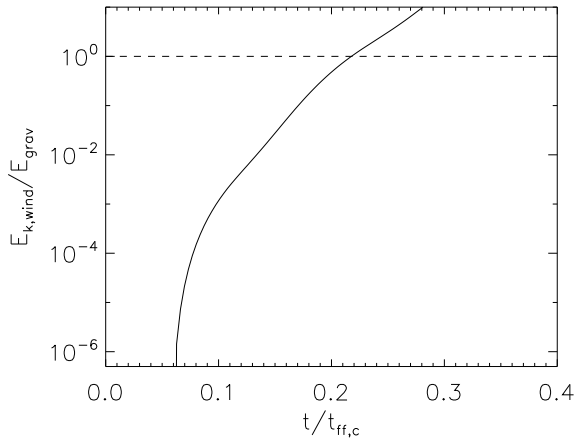,width=\columnwidth}
\end{center}
\caption{Time evolution of the ratio of the kinetic energy generated by stellar winds and the gravitational energy of the protocluster clump. Time is shown in units of the protocluster clump free fall timescale $t_{ff,c}$. The horizontal dashed corresponds to $E_{k,wind}/E_{grav} =1$ with  $\kappa =0.1$.}
\label{fig7}
\end{figure}

\begin{figure}
\begin{center}
\epsfig{figure=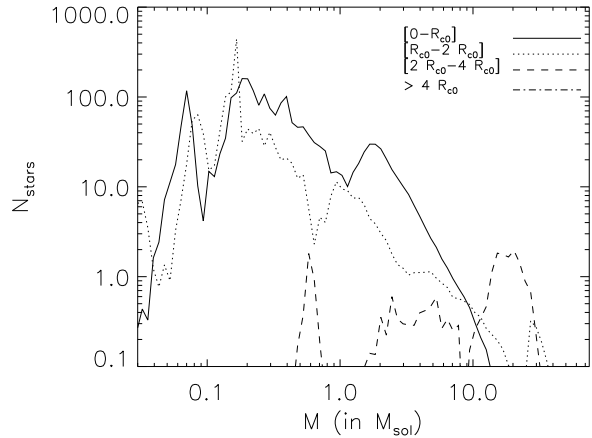,width=\columnwidth}
\end{center}
\caption{The IMF at $t=0.21~t_{ff,c}$in  three different regions of the protocluster clump in the fiducial model: In the central region between the center and the core radius (full line), between one time and two times the core radius (dotted line), between two times and four times the core radius (dash-dotted line), and in fore regions outside four times the core regions (dashed-triple dot line; there is is no stars however in those outer regions). The core radius in this model is $R_{c0}=0.2$ pc. Note that there are no stars present beyond $4~R_{c0}$. The adopted bin size in mass is 2 times bigger than the adopted mass bin in the calculation. This is done to enable a better viewing of the global features and smooth any sharp bin-to-bin variations. The same bin size is adopted in Figs.~\ref{fig10}, \ref{fig12}, \ref{fig14},\ref{fig15}, and \ref{fig17}.}
\label{fig8}
\end{figure}

At the same time, as all the bins of the IMF are populated, at $t \sim 0.21~t_{ff,c}$, the model IMF reproduces almost all of the features of the present day IMF of the ONC. Namely, a shallow slope in the mass range $\sim [0.3-2.5]$ M$_{\odot}$, a steeper slope in the mass range  $\sim [2.5-12]$ M$_{\odot}$, and nearly flat tail at the high mass end. The PSCMF in the model and subsequently the IMF would continue to evolve as time advances. However, at $\sim 0.21~t_{ff,c}$, the kinetic energy generated by stellar winds becomes comparable or larger than the binding gravitational energy in the protocluster clump (see Fig.~\ref{fig7}). This would inevitably lead to the ejection of the gas from the protocluster clump and to the settling of the IMF into a form that is very similar to the present day mass function of the ONC. The dispersal of the gas being dispersed from the central regions is observed with a layer of star forming molecular gas (Genzel \& Stutzki 1989), standing between an outwardly moving ionized gas produced by the central massive stars (O'Dell et al. 1994) and thin layer of neutral gas (O'Dell et al. 1992). 

An important aspect of our model is that it reproduces simultaneously both the shape and the normalization of the present day IMF of the ONC and the mass function of dense sub-millimeter cores in Orion. Note that the synchronization that occurs between the setting of the final IMF and of the gas ejection from the protocluster clumps at $t\sim 0.21 t_{ff,c}$ is related to our choice of $\kappa=0.1$. However, if $\kappa$ had a different value in the range $0.1-0.5$, the epoch of gas dispersal from the cluster would shift to $t \sim 0.18-0.2~t_{ff,c}$, when the model IMF is already in good agreement with that of the ONC. The most massive star in the ONC of mass $\sim 50$ M$_{\odot}$ is not reproduced by the model. This could simply due to very specific conditions in the region of the protocluster clump where this star is born and that are not taken into account by the model. A process that could produce a very massive core from which this massive O star could be born is the coalescence of two or more low mass cores as suggested by Dib et al. (2007a) (see also Edgar \& Clarke 2004). It could otherwise be due to the inadequacy of the accretion rate normalization that we have used in Eq.~\ref{eq20} to describe the accretion process at the very high mass regime, where the effects of radiation from the nascent protostars might be important in regulating the accretion rate from the surrounding medium. Overall, this fiducial model based on the co-evolution of the PSCMF and of the IMF is able to reproduce most of the observed features of both the PSCMF in Orion and the IMF of the ONC. Furthermore, the model suggests that the present day mass function of the ONC has a primordial origin. Thus, the results of our model argue against the dynamical scenario proposed by Pflamm-Altenburg \& Kroupa (2006). 

\subsection{AGE SPREAD OF STARS}\label{age_spread}

Our proposed scenario for star formation in the ONC based on the co-evolution of the PSCMF and of the IMF through gas accretion by the cores, their collapse to form stars and the quenching effect of star formation by stellar winds implies a very small age spread for stars in the ONC. In the fiducial model the first stars are formed at $t  \sim 0.07 t_{ff,c}$ and the star formation is quenched at $t  \sim 0.21 t_{ff,c}$. The age spread in the fiducial model is of the order of $\sim 0.14~t_{ff,c} \sim 0.14 \times 1.62 \times 10^{6}$ yr $\sim 2.3 \times 10^{5}$ yr, where $t_{ff,c} \sim 1.62 \times 10^{6}$ yr is the free-fall time of the protocluster clump with the assumed model parameters. In the ONC, the measured age spread is $\sim 3$ Myr based on age determinations made by fitting isochrones to the ONC's color magnitude diagram for stars with masses $\lesssim 4$ M$_{\odot}$  (Hillenbrand 1997). Palla \& Stahler (1999) argued that the ONC hosts stars that are as old as 10 Myr with the bulk of the stars having an age of $\sim 2$ Myr. However, Hartmann (2003) pointed out that the determination of stellar ages using masses in the range $0.4-6$ M$_{\odot}$, which was also used by Tan el al. (2006) to argue for star formation occurring over several crossing times in the ONC, tend to systematically overestimate the ages of the stars because the birth line age corrections have been underestimated. However, as pointed out by Hillenbrand (1997), about 80 percent of the stars in the ONC have an age that is $\lesssim 0.3$ Myr.  Furthermore, based on their observations of high correlations between the motions of the residual gas and of stars in the ONC, F\"{u}r\'{e}sz et al. (2008) argued that the entire system must be very young with an age of at most one crossing time. Thus, our model reproduces the characteristic age spread of most of the stars in the ONC. The other 20 percent of stars in the ONC with estimated ages $ > 0.3$ Myr may possibly have their ages overestimated,  be foreground stars (Hartmann 2003), or eventually may have formed during the early assembly phase of  the protocluster clump in a scenario similar to the one proposed by Burkert \& Hartmann (2004) and Hartmann \& Burkert (2007).  

\subsection{MASS SEGREGATION}\label{mass_segreg}

\begin{figure}
\begin{center}
\epsfig{figure=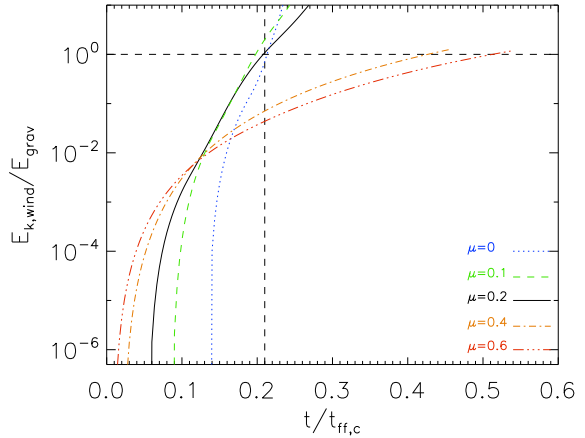,width=\columnwidth}
\end{center}
\caption{Time evolution of the ratio of the wind energy to the gravitational energy in models that have different values of the exponent of the peak density-mass relation $\mu$. The value of $\mu =0.2$ (black line) corresponds to the fiducial model. The vertical line at $t \sim 0.21 t_{ff,c}$ marks the epoch at which the IMFs of the different models are compared. The horizontal dashed line corresponds to $E_{k,wind}/E_{grav} =1$ with  $\kappa =0.1$.}
\label{fig9}
\end{figure}

\begin{figure}
\begin{center}
\epsfig{figure=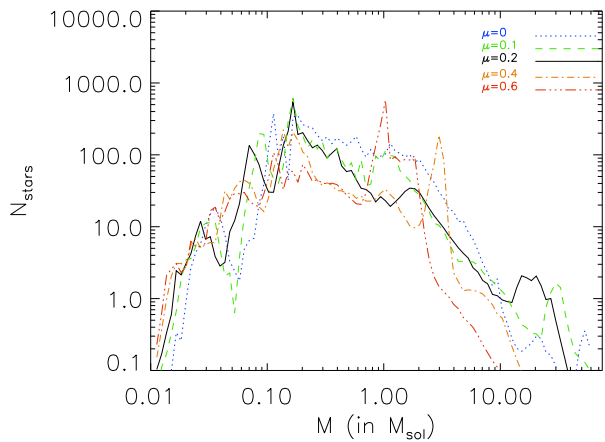,width=\columnwidth}
\end{center}
\caption{The IMF at $t=0.21 t_{ff,c}$ in models which have different values of the exponent of the peak density-mass relation, $\mu$. The value of $\mu =0.2$ (black line) corresponds to the fiducial model.}
\label{fig10}
\end{figure}

Observations of stellar clusters often report evidence for mass segregation with the most massive stars being preferentially located in the inner regions of the cluster (e.g., Pandey et al. 1992; Subramanian et al. 1993; Malumuth \& Heap 1994; Brandl et al. 1996; Hillenbrand \& Hartmann 1998; Fisher et al. 1998; Figer et al. 1999; Sagar et al. 2001; Stolte et al. 2002; Le Duigou \& Kn\"{o}delseder 2002; Sirianni et al. 2002; Lyo et al. 2004; Gouliermis et al. 2004; Sharma et al. 2007,2008). Using N-body simulation with mass-segregated and non-mass segregated clusters, Baumgardt et al. (2008) showed that primordial mass segregation explains better the correlation between the slope of the stellar mass function and the clusters degree of concentration as observed by De Marchi et al. (2007). In the particular case of the ONC, an inspection of the spatial distribution of stars in the cluster (Fig.~3 in Hillenbrand et al. 2007), shows that two out of the six most massive stars are located, in projection, at a distance $r \gtrsim 0.5$ pc from the cluster center, and one star is located, in projection, at a distance of $\gtrsim 1$ pc from the center, whereas the three other stars are located, in projection, within a region $r \lesssim 0.3$ pc from the cluster's center. Bases on time scale estimates Bonnell \& Davies (1998) argued that the massive stars in the ONC must have formed in or near the centre and were not the result of dynamical mass segregation.  

As mentioned in \S~\ref{origin}, the mass segregation of stars in a cluster can have a primordial or a dynamical origin, or both. In this section, we quantify the effects of primordial mass segregation in our accretion-collapse-feedback (ACF) model using the above described fiducial model. Fig.~\ref{fig8} displays the IMF of the fiducial model in different regions of the cluster, namely, in the inner region within one core radius ($R_{c0}=0.2$ pc), in the second annulus between one and two times the core radius, between two and four times the core radius, and for the outer region. Fig.~\ref{fig8} shows that albeit the bulk of the stars with masses $\lesssim 20$ M$_{\odot}$ are found in the inner region (i.e., r  $\lesssim R_{c0}$), the slope of the IMF flattens in the intermediate to high mass regime when going from the inner to the outer regions. The most massive stars are born in the region located between [$2~R_{c0}-4~R_{c0}$]=[$0.4-0.8$] pc. The location of the massive stars in the model is in good agreement with the positions of at least three of the six most massive stars in the ONC. Aside from the fact that one or more of the three stars which are within a projected distance of $\lesssim 0.3$ pc from the ONC's center might be in reality at a larger physical distance from the center in the three-dimensional space, there are also other ways to account for the apparent existence of these three massive stars in the center of the ONC. One possibility is that those three stars have formed, like the other ones, in the region $2~R_{c0}-4~R_{c0}$ and sinked to the center of the cluster by dynamical interactions (e.g., Allison et al. 2009). The second possibility, following ideas developed in Dib et al. (2007a), is that those few stars were born from the coalescence of cores in the center of the proto-ONC clump where the latter are expected to be closely packed and where the coalescence of cores is expected to be more efficient. The coalescence of cores will allow the formation of more massive ones and subsequently, after these cores have collapsed, the formation of the massive stars directly in the central region. The reason why the most massive stars are formed in the region [$2~R_{c0}-4~R_{c0}$]=[0.4-0.8] pc in our model and not in the very inner region lies in the fact that for massive cores of equal masses, the cores located in the very inner region are smaller, as discussed in \S~\ref{core_model}, and thus they have larger average densities and shorter contraction timescales. The shorter contraction timescales of cores in the inner regions will lead them to collapse fatser and thus, have a shorter accretion history which will hinder their development into more massive cores and subsequently more massive stars. Overall, the mild mass segregation observed in the ONC and in our model is most probably another indication of the young age of the cluster. 
 
\section{PARAMETER STUDY}\label{param_study}

\begin{figure}
\begin{center}
\epsfig{figure=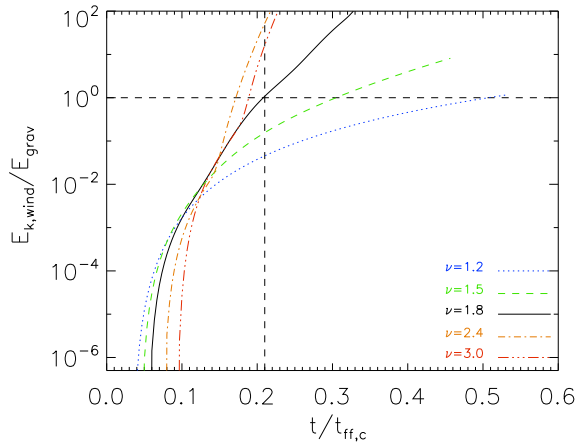,width=\columnwidth}
\end{center}
\caption{Time evolution of the ratio of the wind energy to the gravitational energy in models that have different values of the ratio of the contraction timescale to the free fall time scale of the cores $\nu=t_{cont}/t_{ff}$. The value of $\nu =1.8$ (black line) corresponds to the fiducial model.  The vertical line at $t \sim 0.21~t_{ff,c}$ marks the epoch at which the IMFs of the different models are compared. The horizontal dashed corresponds to $E_{k,wind}/E_{grav} =1$ with $\kappa =0.1$.}
\label{fig11}
\end{figure}

\begin{figure}
\begin{center}
\epsfig{figure=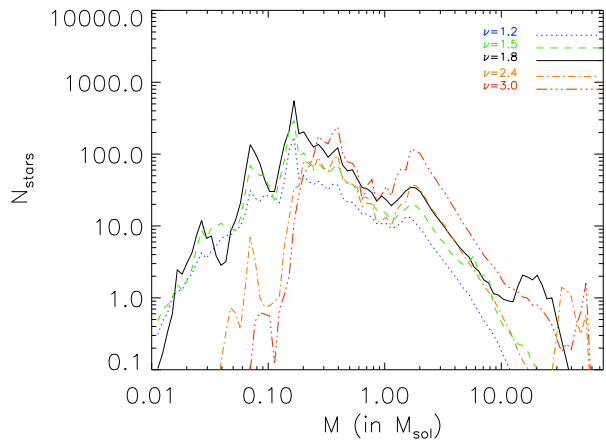,width=\columnwidth}
\end{center}
\caption{The IMF at $t=0.21~t_{ff,c}$ in models where the ratio of the contraction timescale to the free fall time scale of the cores $\nu=t_{cont,p}/t_{ff}=1.2, 1.5,$ and $1.8$ and at $t =0.17~t_{ff,c}$ for $n=2.4$ and $t=0.19~t_{ff,c}$ for $n=3$. The value of $\nu=1.8$ (black line) corresponds to the fiducial model.}
\label{fig12}
\end{figure}

In this section, we investigate the effects on varying some of the model's main parameters that are listed in Tab.~\ref{tab1} on the resulting IMF. However, for the reader conveniency, we first summarize in \S~\ref{summerparam} the parameters of the model and explain their related expected effects.   

\subsection{Summary of the model parameters}\label{summerparam}

In this section, we summarize the free parameters in the model and their range of variations as constrained by observations or theory. We also study the effect of parameter variations on the results of the model. The scheme we adopt to specify the time dependent accretion rate is not affected by free parameters, because it is calibrated using the numerical simulations of SK04. The model contains 8 free parameters listed in Tab.~\ref{tab1}. They can be divided into three categories related to the clump properties, the cores properties, or the feedback model. Of the 8 free parameters in Tab.~\ref{tab1}, the four most important ones are labeled ÒIPÓ,  while the other four are labeled ÒPÓ.
 
In the category of clump parameters, the first one is the clump mass, $M_{c}$. In the fiducial model and in all models below, $M_{c}$ has been assigned the same value of $M_{c}=10^{4} m_{\odot}$, a characteristic mass for protocluster clumps. For a given core formation efficiency, changing $M_{c}$ will in principle only affect the vertical normalization of the PSCMF and of the resulting stellar IMF. However, due to the finite size of the clump and to the $M_{c}-R_{c}$ relation, a more massive clump would have a larger radius, which can introduce some minor effects due to star formation in the outer regions. The second clump parameter is the clump core radius, $R_{c0}$. In all models we chose $R_{c0}=0.2$ pc, which is a characteristic value for stellar clusters and for their protocluster clump progenitors (e.g., Hillenbrand \& Hartmann 1998). The third clump parameter is the exponent of its density profile in the outer regions, $b$. This parameter can be considered to describe the effect of the environment in setting the clump density profile. The most quoted value of $b$ in the literature is $b \sim 2$. However, some observations suggest that $b$ may vary from clump to clump (e.g. Motte et al. 1998). Thus, in the parameter study below, we allow $b$ to vary in the range  $1.4-2.2$, which affects the accretion rate that is directly proportional to the background density. The last of the clump parameters is the rate of core formation efficiency per free fall time, $\epsilon_{c}$. Its value as a function of $M_{c}$ is not well constrained by numerical simulations or theoretical considerations and may vary from clump to clump. Depending on physical conditions, such as the magnetic field strength in the clump, the value of $\epsilon_{c}$ may range from 0.01 or less, to a few times 0.1 (e.g., Nakamura \& Li 2008; V\'{a}zquez-Semadeni et al. 2005b; Padoan \& Nordlund 2009). In all the models we choose $\epsilon_{c}=0.07$. Changing $\epsilon_{c}$ would, to first order, only change the normalization of the PSCMF and of the IMF and not their shapes. This is true as long as the effect of feedback is not considered. If feedback is taken into account, then models with larger $\epsilon_{c}$ values would form a larger number of stars with masses $\gtrsim 10$ M$_{\odot}$ (because of the larger normalization in the total mass of stars) and thus would have a shorter evacuation time of the gas from the protocluster because the wind energy is proportional to $M_{\star}^{4}$.

The two free parameters related to the properties of PSCs are the contraction timescale of the PSCs normalized to their free fall time, $\nu=t_{cont,p}/t_{ff}$, and the exponent of the peak density (or concentration)-mass relation of the PSCs, $\mu$. As will be illustrated below (\S~\ref{contraction}), larger values of $\nu$ imply longer accretion timescales and hence a smaller number of low mass cores. This also implies a relatively larger fraction of intermediate and high mass cores and thus larger fractions of intermediate and high mass stars. As will be shown below in \S~\ref{concentration}, the effect of increasing the value of $\mu$ is to shorten the free-fall time and therefore the accretion timescale. Thus for larger $\mu$ values, PSCs of larger mass will have less time to accrete compared to lower mass PSCs. This results in a decreased fraction of massive stars and a steeper IMF at the high-mass end. 

The last two parameters of the model are those related to the feedback scheme. The mass fraction of the PSC that ends up into a star (the star formation efficiency of an individual PSC), $\xi$, does not affect the shape of the resulting IMF, nor its vertical normalization; it only shifts the IMF horizontally (towards larger or smaller masses). We have assumed $\xi=0.1$ in all models. The parameter $\kappa$ gives the fraction of the winds energy that opposes gravity in the protocluster clump. The value of $\kappa$ would not affect the evolution of the PSCMF and of the IMF if the stellar feedback were not taken into account. The value of $\kappa$ is uncertain and may depend on the other parameters. All we know is that $\kappa \lesssim 1$. The best fit to the ONC data was obtained for $\kappa=0.1$.
 
The radius of the clump, $R_{c}$, and the level of turbulence in the clump described through the exponent of the velocity dispersion-size relation in the clump, $\alpha$, are not free parameters because they are constrained by the observed $M_{c}-R_{c}$ and $R_{c}-v_{c}$ relations (Saito et al. 2007; Eq.~\ref{eq1} and Eq.~\ref{eq2}). Nevertheless, in \S~\ref{effectofturb}, we varied the value of $\alpha$, within a range allowed by the observational uncertainties in those relations, in order to explore the effects of the level of turbulence on the IMF. Higher values of $\alpha$ cause a shift in the position of the characteristic mass in the PN02 model that we adopt as the initial distribution of core masses. 

As the IMF of a stellar cluster is a time evolving distribution until star formation is quenched due to the evacuation of the gas from the protocluster clump, ideally, in investigating the effects of varying the model's main parameters, we would need to compare the IMF of the different models at the time $t=t_{expulse}$, which corresponds to the epoch at which $E_{k,wind}/E_{grav} \sim 1$, when the gas would have been expelled from the protocluster clump. However, as stated above, $\kappa$ is uncertain and may vary from system to system. If a protocluster clump formed massive stars at a slower pace, the condition $E_{k,wind}/E_{grav} \sim 1$ would be reached at a later epoch, leaving more time for other feedback processes not considered in this work, such as jets and outflows, to operate. Furthermore, if massive stars appeared at later stages, our assumption of clump equilibrium may not be justified, and one should perhaps consider the effect of clump contraction. To avoid such complications, we shall compare the IMFs of the different models at the same epoch, independent of whether the gas has been expelled from the protocluster clump, totally or partially. With respect to the values of the parameters in the fiducial model, we vary the values of $b$, $\mu$, $\nu$, and $\alpha$.

\subsection{Effect of the exponent of the mass-concentration relation of the cores}\label{concentration}

Fig.~\ref{fig9} displays the time evolution of the ratio of $E_{k,wind}/E_{grav} $ in five models where $\mu$ has the values of 0, 0.1, 0.2, 0.4, and 0.6. The model corresponding to the case with the fiducial value ($\mu=0.2$) is shown with the black line. The figure shows that the expulsion of the gas occurs at different epochs in the evolution of the protocluster clump for a fixed value of $\kappa$. Fig.~\ref{fig10} displays the IMFs for the different models at $t= 0.21~t_{ff,c}$. The IMF are very similar at the low mass end ($\lesssim 0.1$ M$_{\odot}$) except for a dip in the IMF at $\sim 0.05$ M$_{\odot}$ for the low $\mu$ values. The peak of the IMF is located at $\sim 0.1$ M$_{\odot}$ for the low $\mu$ models and at $\sim 0.8-1$ M$_{\odot}$ in the models with the higher $\mu$ values. The models are also different in the intermediate to high mass end. Low $\mu$ values favor the formation of massive stars and a shallow slope in that mass range, whereas the value of $\psi_{h}$ is much higher in models with higher $\mu$ values. In fact, large values of $\mu$ imply that the cores are more centrally condensed and have smaller contraction timescales than their counterparts in models with lower $\mu$ values. Thus the intermediate and massive cores in the  higher $\mu$ models have shorter accretion timescales than their counterparts with lower $\mu$ values and therefore proceed faster to form stars before more massive cores can be built up by accretion. Thus, larger $\mu$ values, although, they allow massive cores to collapse faster and form a small fraction of massive stars, they delay the formation of a sizable fraction of massive stars and consequently delay the expulsion of the gas from the protocluster clump as can also be seen in Fig.~\ref{fig9}. 

\subsection{Effect of the contraction timescale of the cores}\label{contraction}

Fig.~\ref{fig11} displays the time evolution of the ratio of kinetic energy from stellar winds to the gravitational energy (for $\kappa = 0.1$) for 5 simulations where the ratio of the contraction timescale of the cores to their free-fall time is, at all time during their contraction, maintained at $\nu=1.2, 1.8, 2.4,$ and $3$. Because in the models where the cores contract slower (i.e., higher $\nu$ values, $\nu=2.4,3$), the feedback becomes very important (i.e., $E_{k,wind}/E_{grav} \sim 100$ for $\kappa =0.1$), although massive stars appear slightly later than in models with lower $\nu$ values, the gas is expected to be expelled from the protocluster at an earlier epoch in those models. Thus the IMFs for the different models are compared at $t=0.21~t_{ff,c}$ for the cases where $\nu=1.2, 1.4$, and $1.8$ and at $t=0.17~t_{ff,c}$ for the case with $\nu=2.4$ and at $t=0.19~t_{ff,c}$ for the model with $\nu=3$. The IMFs of the models with various $\nu$ values are displayed in Fig.~\ref{fig12}. The IMFs of models with values of $\nu$ higher than the fiducial value show the presence of massive stars of increasing mass with increasing $\nu$ and a clear absence of low mass stars. On the other hand, the IMFs of models with values of $\nu$ that are lower than the fiducial value show a marked absence of massive stars and similar populations of low mass and a decreasing intermediate mass population with decreasing $\nu$. The accretion history of PSCs with lower $\nu$ values is shorter and inhibits the development of a substantial population of massive cores that would proceed to form massive stars, whereas cores with high $\nu$ values accrete over more extended timescales and evolves towards being more massive with increasing $\nu$ values, thus keeping only a reduced population of low mass core at any given epoch and consequently a small fraction of low mass stars. 

Except for the presence of massive stars, the particular shape of the IMFs in models with high $\nu$ values reminds of the peculiar IMF in Taurus (Luhman et al. 2003b) (the massive stars can disappear from the model IMF if a different normalization is used, i.e., a lower mass clump). The latter exhibits a peak at $\sim 0.8-0.9$ M$_{\odot}$, and a rapid decline when going to lower mass stars. Heyer et al. (2008) showed that the magnetic field in the Taurus molecular cloud complex is relatively strong enough and may cause the cloud to be, at least for the diffuse part, magnetically subcritical. Dense cores that would form in Taurus will thus be likely characterized by longer contraction timescales than in other regions and thus will be able to accrete over more significant periods of time in a similar fashion to the cores with high $\nu$ values in our models.

\subsection{Effect of the protocluster clump density profile}\label{effectofb}

\begin{figure}
\begin{center}
\epsfig{figure=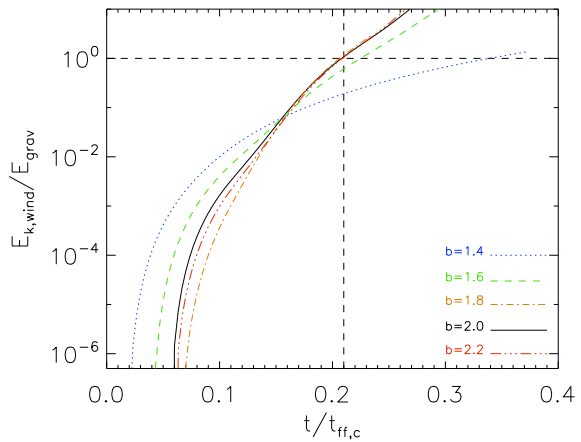,width=\columnwidth}
\end{center}
\caption{Time evolution of the ratio of the wind energy to the gravitational energy in models that have different values of the exponent of the power law profile of the protocluster clumps, $b$. The value of $b=2$ (black line) corresponds to the fiducial model. The vertical line corresponds to the epoch at which $t \sim 0.21~t_{ff,c}$. Note that since the clumps in these models have a different density profile, they consequently have different free-fall times $t_{ff,c}$. The horizontal dashed corresponds to $E_{k,wind}/E_{grav} =1$ with $\kappa =0.1$.}
\label{fig13}
\end{figure}

\begin{figure}
\begin{center}
\epsfig{figure=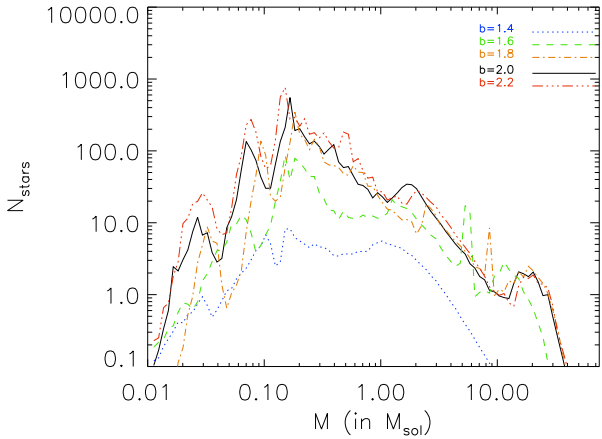,width=\columnwidth}
\end{center}
\caption{The IMF at $t=0 .21~t_{ff,c}$ in models which have different values of the exponent of the power law profile of the protocluster clump, $b$. The value of $b =2$ (black line) corresponds to the fiducial model.}
\label{fig14}
\end{figure}

\begin{figure}
\begin{center}
\epsfig{figure=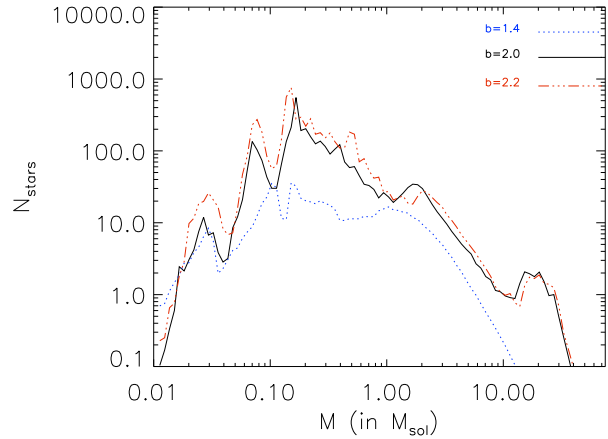,width=\columnwidth}
\end{center}
\caption{The IMF at $t=t_{expell}$ which corresponds to the epoch at which $E_{k,wind/E_{grav} \sim 1}$ (for $\kappa =0.1$) in models which have different values of the exponent of the power law profile of the protocluster clump, $b$. The value of $b =2$ (black line) corresponds to the fiducial model.}
\label{fig15}
\end{figure}

Fig.~\ref{fig13} displays the time evolution of $E_{k,wind}/E_{grav}$ in a series of models where a clump of the same mass as the fiducial model (i.e., $M_{c}=10^{4}$ M$_{\odot}$) possesses a different density profile characterized by the value of $b$ in Eq.~\ref{eq3}. The values of  $b$ that are considered are $b=1.4,1.6,1.8, 2~(fiducial),$ and $2.2$. Thus, for the same value of $M_{c}$, and up to the same radius $R_{c}$, the clumps with higher values of $b$ will be more centrally peaked. The IMFs of the models with different values of $b$ are displayed in Fig.~\ref{fig14}  at $t=0.21~t_{ff,c}$ and in Fig.~\ref{fig15} at $t=t_{expulse}$ which corresponds to the epoch at which $E_{k,wind}/E_{grav} \sim 1$ (with $\kappa$=0.1 in all models).

Whereas the IMF for models for with $b=2$, and $2.2$ display a relatively similar IMF, the IMFs  of models with $b=1.4$ and $b=1.6$ deviate substantially from the IMF of the fiducial model, while the model with $b=1.8$ is intermediate between the two classes. Aside from the issue of normalization (the same core formation efficiency of $\epsilon_{c}= 7$ percent used in all models leads to larger number of cores formed in the more centrally peaked clumps), the slope of the IMF at the high mass end in the models with $b=1.4$ and $b=1.6$ are shallower than in the models with higher $b$ values (alternatively one can say that the plateau in the intermediate mass regime in the low $b$ models is wider). This is due to the fact that in all these models, like in the fiducial one, intermediate mass cores proceed to collapse at a faster pace than their low mass counterparts (because $\mu > 0$, see Fig.~\ref{fig6}). The intermediate mass cores are thus prevented from accreting additional mass and widening the plateau in the mass range $0.1-2$ M$_{\odot}$; whereas lower mass core continue to accrete and become proportionally more numerous in the large $b$ models than in the low $b$ models. When these cores finally proceed to collapse,  the signature of a steeper PSCMF remains imprinted in the generated IMF. If one now compares the IMFs of the models with different $b$ values at $t=t_{expulse}$, the fact that fewer massive stars have formed in the models with low $b$ values delays the evacuation of the gas from the protocluster in those models. Thus, in models with lower $b$ values, additional generations of low and intermediate mass cores have time to form and accrete and cause the widening of the intermediate mass plateau as compared to the case of the high $b$ models.
 
\subsection{Effect of the turbulence level in the protocluster clump}\label{effectofturb}

\begin{figure}
\begin{center}
\epsfig{figure=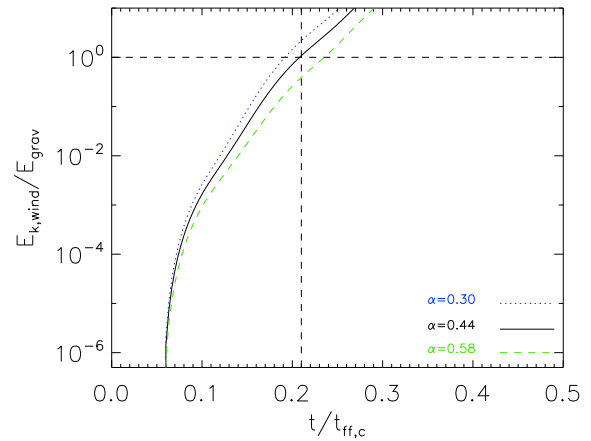,width=\columnwidth}
\end{center}
\caption{Time evolution of the ratio of the wind energy to the gravitational energy in models that have different  exponents of the velocity dispersion size relation in the clump, $\alpha$. The value of $\alpha=0.44$ (black line) corresponds to the fiducial model. The vertical line corresponds to the epoch at which $t \sim 0.21 t_{ff,c}$. The horizontal dashed corresponds to $E_{k,wind}/E_{grav} =1$ with $\kappa =0.1$.}
\label{fig16}
\end{figure}

\begin{figure}
\begin{center}
\epsfig{figure=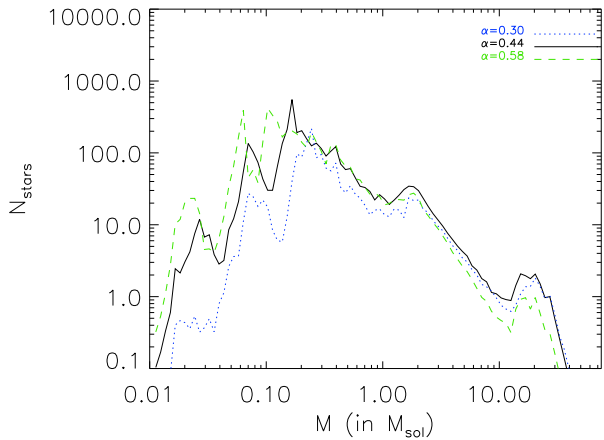,width=\columnwidth}
\end{center}
\caption{The IMF at $t=0.21~t_{ff,c}$ for the models with the exponent of the velocity dispersion-size relation $\alpha=0.44$ (fiducial model) and $\alpha=0.58$ and at $t=0.19~t_{ff,c}$ for $\alpha=0.30$.}
\label{fig17}
\end{figure}

In the previous sections, we have used the velocity dispersion-size relation inside the clumps derived by Saito et al (2007). Here we evaluate the effects on the IMFs of adopting a different velocity dispersion-size relation exponent which takes into account the scatter around the central value in the Saito et al.  data. Since the value  derived by Saito et al. (2007) is $\alpha =0.44 \pm 0.14$, we compare the IMF of the model with the fiducial value, $\alpha=0.44$ to two other models where $\alpha=0.3$ and $\alpha=0.58$. Fig.~\ref{fig16} displays the time evolution of the ratio $E_{k,wind}/E_{grav}$ in the models. Overall, the evolution of this ratio is very similar in the three models, albeit feedback seems to be slightly more important in models with a smaller $\alpha$ value. The corresponding IMFs are displayed in Fig.~\ref{fig17} at $t=0.21~t_{ff,c}$ for models with $\alpha=0.44$, and $\alpha=0.58$, and at $t=0.19~t_{ff,c}$ for the model with $\alpha=0.30$ since for this model the ratio $E_{k,wind}/E_{grav}$ is closer to unity at this epoch. The shapes of the IMFs with different values of $\alpha$ are surprisingly similar at the high mass end. In the absence of time dependent processes such as accretion (which is considered here), the expected slopes of the IMFs would be similar to the slopes of the PSCMF and are expected to have the values of $\psi_{h}=-2.25,-2.41$, and $-2.63$ for $\alpha=0.3,0.44$, and $0.58$, respectively (PN02). The only noticeable difference between these three IMFs and which is due to the effect of varying the velocity dispersion-size relation exponent lies at the low mass end. Higher values of $\alpha$ imply higher Mach numbers and a shift of the characteristic mass in the injected PSCs populations towards smaller masses (PN02) which is the effect that can be observed in Fig.~\ref{fig17}. 

\section{COMPARISON TO PREVIOUS WORK}\label{discuss}

Several authors have investigated the effects of a number of physical processes on the mass function of dense cores or on the IMF (see Bonnell et al. 2007 and references therein). Among those, analytical derivations of the effects of gas accretion on the dense core mass function have been presented by Bonnell et al. (2001),  Basu \& Jones (2004), Bate \& Bonnell (2005), and more recently by Myers (2009). Bonnell et al. (2001b) presented a two stage model for the evolution of the dense core mass function in a protocluster clump. In the Bonnell et al. (2001b) model, in the regime where the gravitational potential of the cluster is dominated by the clump gaseous component, gas is accreted onto a core/star whenever it is located at a distance from it that is smaller to the tidal lobe-radius. As shown by Bonnell et al. (2001b), the result of gas accretion in this regime would be to generate a mass spectrum with a slope of $\psi_{h}=-1.5$. In a second regime where the gravitational potential of the protocluster is dominated by stars, the cores/stars would  accrete following a Bondi-Hoyle like accretion scheme with the accretion radius being equal to the Bondi-Hoyle accretion radius. The result of this accretion is to steepen the slope of the resulting IMF inherited from the previous phase to values in the range of $\sim -2$ to $\sim -2.5$ depending on the initial degree of mass segregation of the stars that have formed so far in the protocluster clump. 

Basu \& Jones (2004) explored the time evolution of a log-normal distribution of the mass function of dense cores under the effect of accretion in a single zone model. They showed that by adopting a mass accretion rate formula of the type $dM/dt \propto M$ (or $dM/dt \propto M^{2/3}$) coupled to a nonuniform accretion timescale of the cores as a function of their mass, the mass function develops a power low tail in the intermediate to high-mass end while retaining its log-normal shape in the low mass regime. Bate \& Bonnell (2005) presented an accretion-ejection scenario for the IMF based on the idea that all objects form with the same initial mass which is set by the choice of the opacity limit, and then accrete at a constant rate until they are ejected from their initial environment. The accretion rate in the Bate \& Bonnell model are drawn from a lognormal distribution of a given mean accretion rate and an associated dispersion. The ejection of the objects is described by a probability exp$(-t/\tau_{eject})$, where $\tau_{eject}$ is a given half life of the ejection process. Bate \& Bonnell (2005) showed that their model can, with appropriate parameters, reproduce some of the standard Galactic field IMFs (i.e., notably the Miller \& Scalo IMF). Myers (2009) showed that the evolution of the dense core mass function under the effect of accretion may also depend on the geometry of the environment in which the spherical star forming clump is embedded. In particular, he showed that the development of a flatter slope at the high mass end is strongly dependent on the environment density and weakly dependent on the environment dimensions.     
  
While our model shares a few conceptual similarities with the above described models, it also contains a number of noticeable different approaches. First, unlike the other models, our prescription for the time dependent accretion rate shape and normalization is parameter free since it is calibrated from numerical simulations. Another difference to the previous models is the multi-zone (various location in the clump) and multi-epoch (cores injected over several epochs) aspect assigned to our treatment of the populations of PSCs and of stars and the co-evolution of two separate populations (only Bonnell et al. 2001b considered a multi-zone model but still not a multi-epoch model). The result of this dichotomy into PSCs and stars is that PSCs that have collapsed to form stars are subtracted from the local PSCs distributions. Also, unlike previous models, feedback from the newly formed stars in taken into account in order to provide an estimate of the epoch at which gas is evacuated from the protocluster environment. The dichotomy into PSCs and stellar populations also allows us to investigate the effect of various core properties on the resulting IMF such as their timescale of contraction (parameter $\nu$) and their mass-peak density relation (parameter $\mu$) which have not been taken into account by any previous model. Obviously the model presented in this paper can be further improved/modified in future works when more accurate theoretical and observational constraints become available. Future improvement might be to adopt an accretion rate prescription that includes the effects of the magnetic field which were not taken into account in the simulations of SK04. The effect of the magnetic field is, in addition to reduce the probability of core formation with increasing field strength (e.g., Dib et al. 2007b), is also to strongly reduce the accretion rates onto the cores (Price \& Bate 2008). Another modification would be to take into account the sub-fragmentation of the cores and build a single-star stellar IMF. As stated in \S~\ref{co_evol}, we made in this work the assumption that a single core leads to a single star/star system and thus the derived IMFs can be compared to system IMFs rather than to single star IMFs. Choosing adapted accretion rates as a function of the degree of magnetization of the clump may also allow us to simultaneously select an adapted value of $\nu$ for a given magnetization level. Further improvements may also include using more constrained values of the parameter $\kappa$ which describes the fraction of the wind energy that  opposes gravity. Simulations of multiple wind interactions are needed to further constrain its value.
  
\section{SUMMARY}\label{summary}

In this work, we use semi-analytical modeling to study the effects of gas accretion onto populations of dense gravitationally bound cores at different locations in a protocluster clump and the transition of the cores to form stars. Once the local contraction timescales for cores of a given mass are shorter than their accretion timescales, cores are turned into stars. We are thus able to follow the co-evolution of the dense core mass function (PSCMF) and of the IMF. Feedback from massive stars in the form of stellar winds is taken into account. The energy from stellar winds that could disperse the gas from the protocluster clump is compared to its gravitational energy and once it becomes dominant, the process of gas accretion onto cores is terminated, and the star formation process is quenched, thus setting up the IMF. We rely on observations and direct numerical simulations in order to calibrate the properties of the protocluster clumps, their populations of cores that are prone to become stars, and the physical processes of gas accretion onto the cores and feedback from massive stars in the form of winds. After prescribing a clump mass, the main free parameters of this Accretion-Collapse-Feedback (ACF) model are a) the exponent of the clump density profile, $b$, b) the contraction timescale of the cores in units of their free-fall timescale $\nu$,  and c) the exponent of the mass-concentration relation of the cores, $\mu$. A fiducial case of the ACF model is discussed in detail and compared to the Orion star forming region. The fiducial case of the ACF model reproduces:

\noindent a) The normalization and complex shape of the IMF of the Orion Nebula Cluster (ONC). Namely, the shallow slope of the IMF of the ONC in the mass range $\sim [0.3-2.5]$ M$_{\odot}$, a steeper slope in the mass range  $\sim [2.5-12]$ M$_{\odot}$, and a nearly flat tail at the high mass end.

\noindent b) The age spread of 80 percent of the young stars in Orion. We find an age spread of $\sim 2.2 \times 10^{5}$, whereas Hillenbrand (1997) suggested that 80 percent of the stars in Orion are younger than $3 \times 10^{5}$ years.   
 
\noindent c) The location of the most massive stars which are born in the region located between $2-4~R_{c0}$ from the clump center, where $R_{c0}=0.2$ pc is the core radius of the clump in the fiducial model.    
 
\noindent d) Simultaneously with the IMF, the normalization and shape and of the mass function of the dense sub-millimeter cores in the star forming regions of Orion which displays a tail at the high mass end. 

In a second step we evaluate the effects of changing some of the models parameters on the resulting IMF. The results are the following:

\noindent e) More concentrated cores with increasing mass (i.e., larger values of $\mu$ in the models), result in steeper IMFs (Fig.~\ref{fig10}), whereas  smaller value of $\mu$ favor the formation of massive stars and the development of a tail-like structure at the high mass end. 

\noindent f)  Longer contraction timescale of the cores (i.e., larger values of $\nu$ in the models) favor a longer and more efficient accretion onto the cores and thus the formation of massive cores and in the same time the depletion of lower mass cores. This results in IMFs which contain an increasing fraction of massive stars and no or very small fractions of smaller mass stars. On the other hand a decreasing contraction timescale of the cores favors the formation of a steeper IMF at the high mass end (Fig.~\ref{fig12}). 

\noindent g) Star formation in a protocluster clump with a radial profile which is close to an $r^{-2}$ show little variations in the resulting IMFs. However, stronger deviations of the density profile of the clump result in significant variations both at the low and high mass end. A steeper radial profile of the clump results in an IMF which contains massive stars and a relatively depleted low mass end, whereas a shallower clump profile results in an IMF which is steeper at the high mass end, shallower at the low mass end, with a broad intermediate mass plateau (Fig.~\ref{fig14}). 

\noindent h) Variations in the level of turbulence in the protocluster clump (i.e., variations in the exponent of the velocity dispersion-size relation) essentially result in variations at the low mass end in the resulting IMFs. Higher levels of turbulence result in an IMF that contains larger fractions of low mass stars (Fig.~\ref{fig17}). The shape of the IMF at the high mass end is unaffected by changing the turbulence levels in the clump. This indicate that the resulting IMF, in this mass regime, is not very sensitive to the exact shape of the mass function of dense cores that are being continuously generated in the clump until star formation is quenched by stellar feedback. 

Overall, it is important to note that the IMF that results from various combinations of the most relevant parameters of the ACF model is usually not well described by a broken power law function or by a log-normal distribution. Instead, the resulting IMFs show complex shapes and genuine variations in all mass regimes  that can extend beyond the statistical uncertainties. Provided there are variations in the structural and dynamical properties of real protoclusters clumps and in the populations of cores prone to star formation within them, our model predicts genuine variations in the stellar IMF.   

\begin{table*}
\centering
\begin{minipage}{15cm}
\begin{tabular}{l l}
\hline
Protocluster clump variables\footnote{The free parameters of the models are marked with (P), and the important ones are marked with (IP). Important parameters are the ones that can cause variation in the shape of the PSCMF and of the IMF whereas the other parameters will only affect the normalization (i.e., vertical shift) and the horizontal shift along the mass axis. Note that $\alpha$ is not a free parameter of the model as it is constrained by the observed $v_{c}-R_{c}$ relation by Saito et al. (2007). However, its value has been varied around the observed central values within the limits of the statistical uncertainties.} & Meaning of the variables \\
\hline
$M_c$~~~~(P)\footnote{In all the models, we chose $M_{c}=10^{4}$ M$_{\odot}$ which is a characteristic mass for protocluster clumps. To first order, changing $M_{c}$ will change the vertical normalization of the PSCMF and of the IMF but not their shapes.} & mass of the clump \\
$R_{c}$ &  radius of the clump \\
$R_{c0}$~~~(P)\footnote{In all models, we chose $R_{c0}=0.2$ which is characteristic of protostellar clumps and of their stellar cluster progenitors (e.g., Hillenbrand \& Hartmann 1998).} & core radius of the clump \\
$v_{c}$ & velocity disperion of the gas in the clump\\
$\rho_{c0}$ & central density of the clump \\
$b$~~~~~~(IP)\footnote{In the observations, $b$ is seen to vary between $\sim 1.5$ to $2.5$. In our models $b$ is varied between $1.4$ and $2.2$.} & exponent of the density profile of the clump at large radii\\
$E_{grav}$ & gravitational energy of the protocluster clump \\
$\alpha$ & exponent of the velocity dispersion-size relation in the clump\\
$\beta$ & exponent of the kinetic energy power spectrum in the clump \\ 
$t_{ff,c}$ & free-fall timescale of the clump \\ 
$\epsilon_{c}$~~~~~(P)\footnote{The core formation efficiency or fraction of the mass of the clump that turn into dense cores per $t_{ff,c}$, $\epsilon_{c}$ may vary from clump to clump depending on physical conditions such as the magnetic field strength in the clump ranging from 1 $\%$ or less, to a few tens of percent (e.g., Nakamura \& Li 2008; V\'{a}zquez-Semadeni et al. 2005b ;Padoan \& Nordlund 2009). In all the models $\epsilon_{c}$ was set equal to 0.07. Note that similarly to $M_{c}$, the choice of $\epsilon_{c}$, as long as feedback is not considered, does not change the shapes of the PSCMF and of the IMF but only their vertical normalization.} & fraction of the mass of the clump that turn into dense cores per $t_{ff,c}$ \\
\hline
Core variables & Meaning of the variables \\
\hline
 $M$ & mass of the core \\
 $\rho_{p0}$ & central density of the core \\
$\mu$~~~~~(IP)\footnote{In our models, the exponent of the $\rho_{p0}-M$ relation of the cores has been allowed  to vary in the range 0-0.6 in agreement with the range determined from observations (e.g.,  Caselli \& Myers 1995; Johnstone \& Bally 2006).} & exponent of the $\rho_{p0}-M$ relation of the core \\
$R_{p0}$ & core radius of the core  \\
$R_{p}$ & radius of the core \\
$\bar{\rho_{p}}$ & average density of the core \\ 
$t_{ff}$ & free-fall timescale of the core \\
$t_{cont,p}$ & contraction timescale of the core \\
$\nu$~~~~~~(IP)\footnote{The value of $\nu$ is expected to vary between 1 and 10. Numerical and observational determinations of $\nu$ suggest that it is in the range of a few (e.g., Ward-Thompson et al. 2007; Galv\'{a}n-Madrid et al. 2007).} & ratio of the contraction timescale of the core to its free-fall time \\
\hline
Accretion variables & Meaning of the variables \\
 \hline
 $\dot{M}_{acc,eff}$ & effective accretion rate of the core \\
\hline
Stellar and Feedback variable & Meaning of the variables \\ 
\hline
$\xi$~~~~~~(P)\footnote{In all models, $\xi$ has been fixed to the value of $0.1$. Varying $\xi$ does not change the shape of the resulting IMF but only shifts it horizontally along the mass axis.} & fraction of the mass of the core that ends up in the stars\\
$M_{\star}$ & masses of stars \\
$\dot{M}_{\star}$ & mass loss rates from stars \\ 
$E_{wind}$ & time integrated energy from massive stars for $M \ge 10$ M$_{\odot}$  \\
$\kappa$~~~~~~(IP)\footnote{In all models, $\kappa$ has been fixed to the value of $0.1$. As explained in the text, all that is presently known is that $\kappa \leq 1$. Simulations of multiple wind interactions are needed to further constrain its value.} & fraction of $E_{wind}$ converted into motions that oppose the gravity  of the clump \\  
$E_{k,wind}$ & time integrated energy from stellar winds that oppose the gravity of the clump \\
\hline
 \end{tabular}
\end{minipage}
\caption{Main variables in the Accretion-Collapse-Feedback model. From top to bottom, the first panel describes the protocluster clump variables, the second panel the dense cores variables, the third panel the acrretion variable, and the lower panel the stellar and stellar feedback variables.}
\label{tab1}
\end{table*}

\section*{Acknowledgments}

We would like to thank the referee for many useful comments and suggestions which have lead to a much improved version of the paper. We are also very grateful to Stefan Schmeja, Ralf Klessen, Hiro Saito, Edouard Audit, Patrick Hennebelle, Gilles Chabrier, Shantanu Basu, Fr\'{e}d\'{e}rique Motte, Philippe Andr\'{e}, David Nutter, Luis Felipe Rodriguez, Vasilii Gvaramadze, Andrea Stolte, Sungsoo Kim, and Francesco Miniati for very interesting discussions and feedback regarding this work. S. D. is supported by the project MAGNET, financed by the Agence Nationale de la Recherche (France) and aknowledges the financial support from the Indo-French Astronomy Network (IFAN) and the hospitality of the TIFR (Mumbai), IUCAA (Pune), and IIA (Bangalore). F. K. acknowledges the support of an Ad Astra PhD scholarship from the University College Dublin.
 
{}

\label{lastpage}

\end{document}